\documentclass[prc,amsfonts,reprint, superscriptaddress,showpacs,onecolumn, notitlepage,nofootinbib,longbibliography,11pt]{revtex4-1} 
\pdfoutput=1
\usepackage{amssymb}
\usepackage{amsmath}
\usepackage{color}
\usepackage{graphicx}
\usepackage{soul}

\newcommand{\comment}[1]{}

\begin{document}

\title{Transmission spectra of an ultrastrongly coupled qubit-dissipative resonator system}

\author{L. Magazz\`u}
\affiliation{Institute for Theoretical Physics, University of
Regensburg, 93040 Regensburg, Germany}

\author{M. Grifoni}
\affiliation{Institute for Theoretical Physics, University of Regensburg, 93040 Regensburg, Germany}

\date{\today}

\begin{abstract}
We calculate the transmission spectra of a flux qubit coupled to a dissipative resonator in the ultrastrong coupling regime. Such a qubit-oscillator system constitutes the building block of superconducting circuit QED platforms. The calculated transmission of a weak probe field quantifies the response of the qubit, in  frequency domain, under the sole influence of the oscillator and of its dissipative environment, an Ohmic heat bath. We find the distinctive features of the qubit-resonator system, namely  two-dip structures in the calculated transmission, modified by the presence of the dissipative environment. The relative magnitude, positions, and broadening of the dips are determined by the interplay among qubit-oscillator detuning, the strength of their coupling, and the interaction with the heat bath. 
\end{abstract}

\maketitle

\section{Introduction}
Current developments in circuit quantum electrodynamics (QED) are establishing superconducting devices as leading platforms for quantum information and simulations ~\cite{You2011,Koch2012,Pekola2015,Wendin2017, Nori2017}. In particular, quantum optics experiments with qubit coupled to superconducting resonators are now performed in (and beyond) the so-called ultrastrong coupling (USC) regime, with the qubit-resonator coupling reaching the same order of magnitude of the qubit splitting and resonator frequency~\cite{Niemczyk2010,Ashhab2010,Forn-Diaz2010,Yoshihara2017, Yoshihara2017PRA,Forn-Diaz2018review,Kockum2019}.
The qubits are essentially based on superconducting circuits interrupted by Josephson junctions, the nonlinear elements that provide the anharmonicity required to single-out the two lowest energy states~\cite{Devoret2004}. In the flux configuration, the qubit states are superpositions of the eigenstates of the magnetic flux operator associated to   clockwise and anti-clockwise circulating supercurrents, corresponding to the two lowest energy eigenstates of a double-well potential \emph{seen} by the flux coordinate.
The double-well can be biased by applying an external magnetic flux and transitions between states in this qubit basis, where the states are localized in the wells, occur via tunneling through the potential barrier of the potential.\\
\indent The standard theoretical tool to account for the coupling of superconducting qubits to their electromagnetic or phononic environments is provided by the spin-boson model (SBM), consisting of a quantum two-level system interacting with a heat bath of harmonic oscillators~\cite{Leggett1987,Weiss2012}. This model has been the subject of extensive studies as an archetype of dissipation in quantum mechanics and the different coupling regimes of spin-boson systems and the associated dynamical behaviors have been theoretically explored by using a variety of approaches~\cite{Petruccione2002,Weiss2012}.
Only recently though, progress in the design of superconducting circuits have opened the possibility to attain experimental control on the strong qubit-environment coupling regime ~\cite{Forn-Diaz2017,Magazzu2018,Ustinov2018,Roch2019, Kuzmin2019}.\\
\indent In circuit QED, an appropriate description for qubit-resonator systems is provided by the Rabi Hamiltonain, whose interaction part features energy-nonconserving terms called counter-rotating. In this context,  USC refers to an interaction regime where the rotating wave approximation, that allows for a description in terms of the Jaynes-Cummings Hamiltonian, appropriate for atom-cavity systems, fails, as the counter-rotating terms cannot be neglected~\cite{Forn-Diaz2018review,Kockum2019}. A refined classification of the different regimes of the Rabi model is provided in~\cite{Rossatto2017}. The USC regime of circuit QED is currently the subject of much theoretical work, see for example~\cite{Garziano2015,DiazCamacho2016, Armata2017,DeBernardis2018,DiStefano2019}.\\
\indent In the present work we consider a flux qubit ultrastrongly coupled to a superconducting resonator, modeled as a harmonic oscillator, which in turn  interacts with a bosonic heat bath.  The qubit is probed by an incoming field whose transmitted part provides information on the dynamics under the influence of the  resonator and its environment. While weak dissipation affecting a USC system as a whole has been addressed via a master equation approach in~\cite{Blais2011}, here we consider the case where the coupling to the environment, of arbitrary strength, affects the resonator exclusively. The setup considered  describes quantum optics experiments in circuit QED  but also the coupling of a qubit to a detector~\cite{Chiorescu2004,Thorwart2004, Johansson2006} and the qubit-bath coupling mediated by a waveguide resonator in a heat transport platform in the  quantum regime~\cite{Pekola2018}. Alternatives to the spectroscopy of the qubit to investigate USC systems exist. For example, spectroscopy of ancillary qubits has been proposed in~\cite{Lolli2015} to probe the ground states of ultrastrongly-coupled systems. Moreover, methods alternative to the analysis of the transmission spectra have been recently devised to probe the USC regime~\cite{Falci2019,Ridolfo2019}.\\ 
\indent The dissipative Rabi model which describes our setup can be mapped to a SBM where the spin interacts directly with a bosonic bath characterized by an effective spectral density function peaked at the oscillator frequency~\cite{Garg1985}, which constitutes a so-called structured environment. Using the same approach as the one developed in~\cite{Magazzu2018} to analyze the  measured transmission of a probe field in the presence of a Ohmic environment and of a pump drive, here we calculate the transmission spectra of the qubit, considering different qubit-resonator detuning and coupling strengths. By employing a nonperturbative approach to include the dissipation, we find that the characteristic two-dip profiles of the transmission are affected nontrivially by the presence of the bath beyond the weak dissipation limit.
 The picture in which position and relative magnitude of the dips are determined by the qubit-resonator coupling strength and detuning is modified by bath-induced renormalization effects. In particular, the renormalization of the resonator frequency affects the relevant transition frequencies in a non-symmetric way, reducing the so-called vacuum Rabi splitting. At large resonator frequencies, an Ohmic-like qubit transmission is recovered whereas, for low frequencies of the resonator, the single,  broadened dip in the transmission displays an upwards renormalization of the qubit splitting.

 \section{Flux qubit coupled to a  dissipative resonator}
\label{setup}
 The model for a time-dependent open system coupled to an environment of mutually independent bosonic modes, with possible partitioning into sub-environments  (heat baths), is provided by the Hamiltonian
\begin{equation}\label{Hgeneral}
H(t)=H_{\rm S}(t)+\hat{A}_{\rm S}\sum_k \lambda_k (a_k^{\dagger}+a_k)+\sum_k\hbar\omega_k a_k^{\dagger}a_k\;,
\end{equation}
where $\hat{A}_{\rm S}$ is a system operator and where the bath operators $a_k^\dag$ and $a_k$ create and destroy, respectively, an excitation in the $k$-th harmonic oscillator. The angular frequency $\lambda_k$ is the coupling strength between the qubit and the $k$-th harmonic oscillator.
 The bath is fully characterized by the spectral density function 
\begin{equation}\begin{aligned}\label{}
G(\omega)=\sum_k\lambda_k\delta(\omega-\omega_k)\;.
\end{aligned}\end{equation}
In the continuum limit $G(\omega)$ is usually taken to be proportional  to a power of $\omega$ at low frequencies and to have a cutoff at high frequencies. Moreover, the overall coupling to the bath is quantified by a single parameter $\alpha$. A prominent example is the Ohmic bath for which $G(\omega)=2\alpha\omega f_{\rm c}(\omega)$, where $ f_{\rm c}(\omega)$ is a cutoff function.\\
\begin{figure}[ht!]
\begin{center}
\includegraphics[width=0.9\textwidth,angle=0]{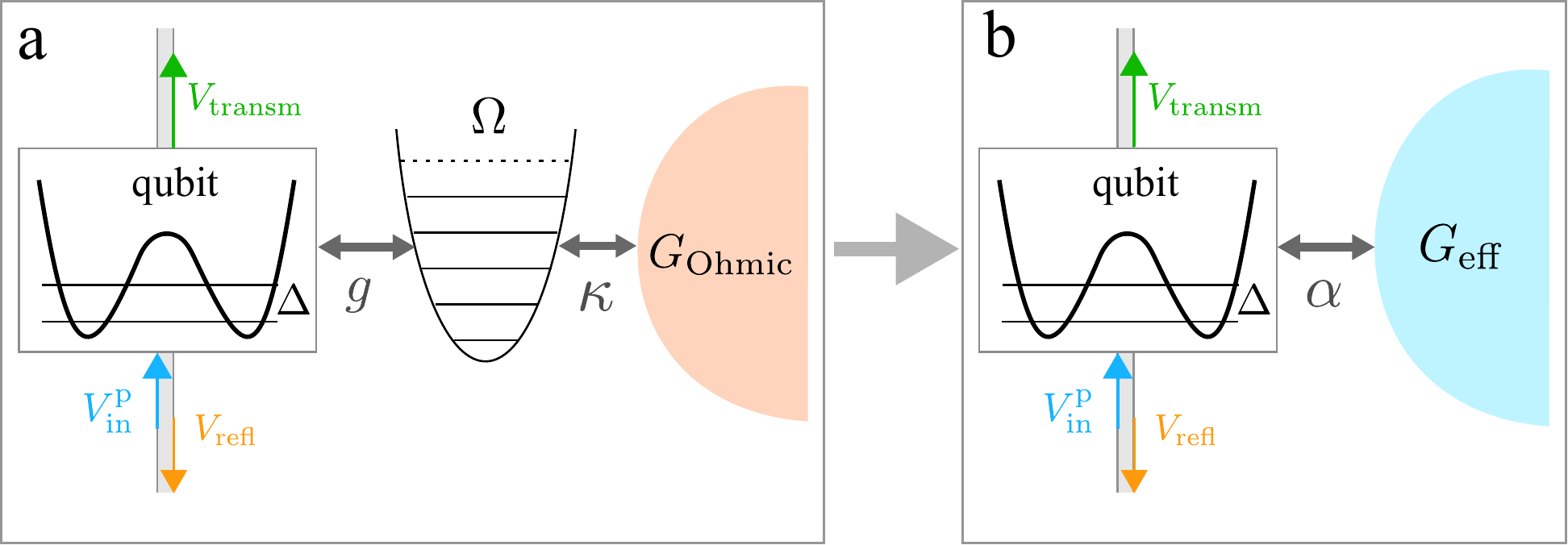}
\caption{\small{ \textbf{a} - Scheme of the setup analyzed. A flux qubit, probed through a transmission line, is coupled to a resonator, the harmonic oscillator of frequency $\Omega$. The latter is in turn in contact with a Ohmic heat bath. The incoming probe field $V_{\rm in}^{\rm p}$ is scattered at the qubit position, resulting in a transmitted and a reflected field. \textbf{b} - Mapping to the spin-boson model. The harmonic bath is described by the structured effective spectral density of Eq.~\eqref{Geff}, with effective coupling $\alpha$.}}
\label{scheme}
\end{center}
\end{figure}
\indent  In the present work we consider a qubit-resonator system, with the resonator coupled to a bosonic heat bath according to the scheme in Fig.~\ref{scheme}-a. The qubit is characterized by the frequency scale  $\Delta$. The resonator is modeled as a harmonic oscillator of frequency $\Omega$ and the frequency $g$ is the qubit-resonator coupling.  The resonator is in contact with a dissipative environment modeled as a strictly Ohmic bath with spectral density $G_{\rm Ohmic}(\omega)=\kappa \omega$, where the dimensionless parameter  $\kappa$ quantifies the overall oscillator-bath coupling.
 The resulting system is described by the dissipative Rabi Hamiltonian, namely by Eq.~\eqref{Hgeneral} with $H_{\rm S}(t)\equiv H_{\rm Rabi}(t)$, where
\begin{equation}\begin{aligned}\label{HRabi}
H_{\rm Rabi}(t)=-\frac{\hbar}{2}\left[\Delta\sigma_x+\varepsilon(t)\sigma_z\right]+\hbar \Omega B^{\dagger}B+\hbar g\sigma_z(B^{\dagger}+B)\;,
\end{aligned}\end{equation}
and with $\hat{A}_{\rm S}=\hbar(B^{\dagger}+B)$. 
Here, $B^\dag$ and $B$ are the resonator mode operators and the operators $\sigma_j$ are the Pauli spin operators in the qubit  basis. The qubit parameters, with dimensions of an angular frequency, are the time-dependent bias $\varepsilon(t)$ and the  bare qubit frequency splitting at zero bias $\Delta$. In a truncated double-well potential realization of the two-level system, which is proper of flux qubits, $\Delta$ is the tunneling amplitude per unit time of the isolated qubit. 
The qubit bias $\varepsilon(t)$ is (weakly) driven by an incoming probe field through a transmission line which is an independent part of the setup.\\
\indent Within Van Vleck perturbation theory~\cite{VanVleck1929,CohenTannoudji1998}, with $g$ treated as a small parameter with respect to $\Delta$ and $\Omega$, the spectrum of the Rabi Hamiltonian in Eq.~\eqref{HRabi}, can be calculated analytically~\cite{Hausinger2008}. In the unbiased case, $\varepsilon(t)=0$, the eigenfrequencies of the ground state and of the first two excited states read
\begin{equation}\begin{aligned}\label{Rabi_spectrum}
\omega_0=&-\frac{\Delta}{2}- f(\Omega)\\
\omega_{1/2}=&\frac{\Omega}{2}- f(\Omega)\mp \frac{1}{2}\sqrt{[\Delta-\Omega+2f(\Omega)]^2+4g^2}\;,
\end{aligned}\end{equation}
with $f(\Omega)=g^2\Delta^2[\Delta^2(\Delta+\Omega)]^{-1}$. For $\Omega=\Delta$, i.e. at zero detuning, the spectrum presents avoided crossings, see Fig.~\ref{colormap}-b below, and the difference $\omega_2-\omega_1\simeq 2 g$ is the so-called vacuum Rabi splitting, see~\cite{Wallraff2004,Wallraff2008} for experimental observations. Van Vleck perturbation theory can also be used at arbirtary coupling, and thus in the USC regime, also in the presence of a static bias, by treating the qubit energy splitting as a small parameter~\cite{Hausinger2010PRA}.\\
\indent The dissipative Rabi Hamiltonian given by Eqs.~\eqref{Hgeneral} and~\eqref{HRabi} can be mapped, see Fig.~\ref{scheme}-b, to the Hamiltonian of the SBM~\cite{Grifoni1998,Weiss2012}
\begin{equation}\label{HSBM}
H_{\rm SB}(t)=-\frac{\hbar}{2}\left[\Delta\sigma_x+\varepsilon(t)\sigma_z\right]-\frac{\hbar}{2}\sigma_z\sum_k \lambda_k (a_k^{\dagger}+a_k)+\sum_k\hbar\omega_k a_k^{\dagger}a_k\;,
\end{equation}
where the qubit is directly coupled to a structured bosonic bath with an effective spectral density function that, in the continuum limit, reads~\cite{Garg1985,Goorden2004,Zueco2019}
\begin{equation}
\label{Geff}
G_{\rm eff}(\omega)=  \frac{2 \alpha \omega
\Omega^4}{(\Omega^2-\omega^2)^2+(2\Gamma\omega)^2}\;.
\end{equation}
This effective spectral density function has an  Ohmic behavior ($\propto\omega$) at low frequencies, $\omega/\Omega\ll 1$, and features a Lorentzian peak centered at the oscillator frequency $\Omega$  with semi-width $\Gamma=\pi\kappa\Omega$, with $\kappa$ the oscillator-bath coupling strength. The effective coupling strength between the qubit and the structured bath is given by the dimensionless parameter $\alpha=8 \kappa g^2/\Omega^2$. Note that for large $\Omega$ the Ohmic case with weak coupling is recovered from the spectral density in Eq.~\eqref{Geff}. Such mapping can be seen as the inverse application of the reaction coordinate mapping, a technique used to deal with open systems in structured environments~\cite{Martinazzo2011}. 

\section{The driven spin-boson model within NIBA}\label{sec_driven_NIBA}
The exact time evolution of the qubit population difference $P(t)=\langle \sigma_z(t)\rangle$ in the SBM is governed by the generalized master equation (GME)~\cite{Grifoni1998}
\begin{eqnarray}\label{GME}
\dot{P}(t)=\int_{t_0}^tdt'\left[ \mathcal{K}^{-}(t,t')-\mathcal{K}^{+}(t,t')P(t')\right]\;.
\end{eqnarray}
The formal exact expression for the kernels can be found within the path integral representation of the qubit reduced density matrix. In the path integral approach, the Feynman-Vernon influence functional~\cite{Feynman1963}, which results from  tracing out exactly the environmental degrees of freedom, couples the qubit tunneling transitions in a time-nonlocal fashion. The exact kernels of the GME collect all the irreducible sequences of tunneling processes involved in the sum over paths, namely the sequences that cannot be cut into two or more noninteracting parts, the interactions being mediated by the bath correlation function $Q(t)$
\begin{equation}\label{Qdef}
Q(t)= Q'(t)+{\rm i} Q''(t)=\int_{0}^{\infty}d\omega \frac{G(\omega)}{\omega^2}\left[ \coth{\left(\frac{\hbar\omega\beta_\nu}{2}\right)}(1-\cos{\omega t})+{\rm i}\sin{\omega t} \right]\;.
\end{equation}
This function is related to the bath force operator, or quantum noise,
\begin{equation}\label{zeta}
\hat{\xi}(t)=\sum_{j=1}^Nc_{j}\left[\hat{x}_{j}(0)\cos{(\omega_{j} t)}+\frac{\hat{p}_{j}(0)}{m_{j}\omega_{j}}\sin{(\omega_{j} t)}\right]
\end{equation}
which appears in the generalized quantum Langevin equation~\cite{Hanggi2005} for a central system coupled to the harmonic environment described in Eq. (\ref{Hgeneral}). Here  $\hat{x}_j$ and $\hat{p}_j$ are position and momentum of the $j$-th bath oscillator. The bath correlation function $Q(t)$ coincides with the two-time integrated  bath force correlation function $L(t)=\langle \hat{\xi}(t)\hat{\xi}(0)\rangle$, i.e., $\ddot{Q}(t)=L(t)$. \\
\indent The exact formal expression for the kernels of the GME has no known closed form that can be used for actual calculations, so that approximation schemes appropriate for the different  physical parameter regimes are introduced. The  noninteracting-blip approximation (NIBA) exploits the fact that the time-nonlocal interactions, the so-called blip-blip interactions mediated by $Q(t)$, are suppressed at long times and to an extent that increases with the coupling strength to the heat bath and with the bath temperature. This approximation scheme consists in neglecting these nonlocal interactions and is therefore suited for the strong coupling/high temperature regime.  
However, at zero bias, $\varepsilon(t)=0$, an exact cancellation of the blip-blip interactions occurs, so that NIBA yields reasonably accurate results for the population difference also at weak coupling~\cite{Nesi2007,Weiss2012}.\\  
\indent The time dependent bias $\varepsilon(t)$ in the qubit Hamiltonian, see Eq.~\eqref{HRabi}, accounts for an externally applied static flux and the monochromatic, weak probe field. Following~\cite{Magazzu2018}, we set
\begin{equation}\begin{aligned}\label{bias_t}
\varepsilon(t)=\varepsilon_0 + \varepsilon_{\rm p}\cos(\omega_{\rm p}t)\;,
\end{aligned}\end{equation}
with $\varepsilon_{\rm p}/\omega_{\rm p}\ll 1$. 
Note that in the actual setup considered in the application of Sec. \ref{results}, the qubit is not biased, meaning that the applied static flux is tuned so as to have $\varepsilon_0=0$.\\
\indent In the presence of the time-dipenent bias in Eq.~\eqref{bias_t}, the kernels of the GME~\eqref{GME}, within NIBA, read~\cite{Grifoni1998}  
\begin{equation}\begin{aligned}\label{K}
\mathcal{K}^{+}(t,t')=& h^{+}(t-t')\cos\left[\zeta(t,t')\right]\;,\\
\mathcal{K}^{-}(t,t')=& h^{-}(t-t')\sin\left[\zeta(t,t')\right]\;,
\end{aligned}\end{equation}
%
where the dynamical phase $\zeta(t,t')$ is given by
\begin{equation}\begin{aligned}\label{zeta}
\zeta(t,t')=&\int_{t'}^{t}dt''\;\varepsilon(t'')\\
=&\varepsilon_0 (t-t')+\frac{\varepsilon_{\rm p}}{\omega_{\rm p}}\left[\sin(\omega_{\rm p} t)-\sin\left(\omega_{\rm p} t' \right)\right]
\;,
\end{aligned}\end{equation}
and where 
\begin{equation}\begin{aligned}\label{hpm}
h^{+}(t)=& \Delta^2 e^{-Q'(t)}\cos[Q''(t)]\;,\\
h^{-}(t)=& \Delta^2 e^{-Q'(t)}\sin[Q''(t)]\;.
\end{aligned}\end{equation}

\section{Spectroscopy of the qubit: Relating the measured  transmission to the qubit dynamics}
\label{section_transmission}
As shown in Fig.~\ref{scheme} (see also~\cite{Peropadre2013,Magazzu2018}), a probe voltage field $V_{\rm in}^{\rm p}(t)=f_{\rm Z}\varepsilon_{\rm p}\cos(\omega_{\rm p} t)$ is scattered by the qubit placed at the center of the transmission line used to probe the qubit.  The constant $f_{\rm Z}$ has dimensions of a flux and the angular frequency $\varepsilon_{\rm p}$ is the (small) amplitude of the probe. The scattering of the incoming probe field $V_{\rm in}^p$ yields a transmitted and a reflected field, denoted by $V_{\rm transm}(t)$ and $V_{\rm refl}(t)$, respectively, see Fig.~\ref{scheme}. The flux difference across the qubit is $\delta\Phi(t)=\Phi(0^-,t)-\Phi(0^+,t)$, with the flux given by $\Phi(0^\pm,t)=\int_{-\infty}^t dt'\;V(0^\pm,t')$. Here $0^\pm$ refers to the positions immediately before and after the position of the qubit in the transmission line.\\
\indent Following~\cite{Vool2017}, we have for the voltage  $V(0^-,t)$ and current $I(0^-,t)$ immediately before the qubit position the following equations
\begin{eqnarray}
V(0^-,t)&=&V_{\rm in}^{\rm p}(t)+V_{\rm refl}(t)\;,\label{SLfields-a}\\
I(0^-,t)&=&\frac{1}{Z}\left[V_{\rm in}^{\rm p}(t)-V_{\rm refl}(t)\right]\;,\label{SLfields-b}
\end{eqnarray}
where $Z$ is the characteristic impedance of the transmission line. Similarly, immediately after the qubit position
\begin{eqnarray}
V(0^+,t)&=&V_{\rm transm}(t)\;,\label{SRfields-a}\\
I(0^+,t)&=&\frac{1}{Z}V_{\rm transm}(t)\;.\label{SRfields-b}
\end{eqnarray}
Using the conservation of the current, $I(0^-,t)=I(0^+,t)$, and the relation $V(0^-,t)-V(0^+,t)=\dot{\delta\Phi}(t)$, we obtain
\begin{eqnarray}\label{Vtransm}
V_{\rm transm}(t)&=&V_{\rm in}^{\rm p}(t)-\frac{\dot{\delta\Phi}(t)}{2}\;.
\end{eqnarray}
The connection between the measured transmission, defined as the ratio
\begin{equation}\begin{aligned}\label{T}
\mathcal{T}(\omega_{\rm p})=V_{\rm transm}(\omega_{\rm p})/V_{\rm in}^{\rm p}(\omega_{\rm p})\;,
\end{aligned}\end{equation}
and the spin-boson dynamics is completed by identifying the flux difference across the qubit with the population difference of the localized eigenstates of the flux operator $\hat{\Phi}=f\sigma_z$, namely by setting $\delta\Phi(t)\equiv f\langle\sigma_z(t)\rangle$, with $f$ a proportionality  constant with dimension of a flux.\\
\indent Since we want to connect the asymptotic, time-periodic dynamics induced by the probe field -- and rendered by the GME~\eqref{GME} -- to the measured transmission, we start by considering the asymptotic population difference $P_\infty(t)=\lim_{t\to\infty}P(t)$. Due to its periodicity, with period $2\pi/\omega_{\rm p}$, we can express it as a Fourier series whose time derivative reads  
\begin{eqnarray}\label{Pasdot}
\dot{P}_\infty(t)&=&\sum_m {\rm i}m\omega_{\rm p} p_m e^{{\rm i}m\omega_{\rm p} t}\;,
\end{eqnarray}
where 
\begin{eqnarray}\label{Scoeff}
p_m=\frac{\omega_{\rm p}}{2\pi}\int_{-\mathcal{\pi}/\omega_{\rm p}}^{\mathcal{\pi}/\omega_{\rm p}}dt\; P^{\rm as}(t)e^{-{\rm i}m\omega_{\rm p} t}\;.
\end{eqnarray}
In the asymptotic limit we set $\delta\Phi(t)\equiv fP_\infty (t)$. Then, the transmission at the probe frequency $\omega_{\rm p}$ ($m=1$ in Eq.~\eqref{Pasdot}) obtained by plugging Eq.~\eqref{Vtransm} into Eq.~\eqref{T}, is given by
\begin{equation}\begin{aligned}\label{Transmission}
\mathcal T(\omega_{\rm p})=&\frac{f_{\rm Z}\varepsilon_{\rm p}/2-{\rm i}f\omega_{\rm p} p_1/2}{f_{\rm Z}\varepsilon_{\rm p}/2}\\
=&1-{\rm i}\mathcal{N}\omega_{\rm p}  p_1/\varepsilon_{\rm p}\;,
\end{aligned}\end{equation}
where $\mathcal{N}= f/f_{\rm Z}$. The parameter $\mathcal{N}$ can be estimated in experiments and will be set to an arbitrary value in the application of Sec. \ref{results}. \\
\indent Due to the effect of the monochromatic probe, the asymptotic population $P_\infty(t)$ is periodic with the period of the probe. Moreover, since the probe is weak ($\varepsilon_{\rm p}/\omega_{\rm p}\ll 1$), we can confine ourselves to the linear response regime, which 
amounts to neglecting terms of order higher than the first in the ratio $\varepsilon_{\rm p}/\omega_{\rm p}$ in the series for $P_\infty(t)$. Denoting with $^{(1)}$ the first order, we obtain~\cite{Grifoni1995,Grifoni1998}
\begin{eqnarray}\label{Pas}
P_\infty(t)&\simeq&p_0+p_1^{(1)} e^{{\rm i}\omega_{\rm p} t}+p_{-1}^{(1)} e^{-{\rm i}\omega_{\rm p} t}\nonumber\\
&\equiv&P_0+\hbar \varepsilon_{\rm  p}[\chi(\omega_{\rm p})e^{{\rm i}\omega_{\rm p}t}+\chi(-\omega_{\rm p})e^{-{\rm i}\omega_{\rm p}t}]\;,
\end{eqnarray}
where we have introduced the linear susceptibility 
\begin{equation}\begin{aligned}\label{chi}
\chi(\omega)=p_1^{(1)}
/\hbar\varepsilon_{\rm p}
\end{aligned}\end{equation}
and where $P_0$ is the equilibrium value of $P(t)$ in the static system. We can then relate the transmission at the probe frequency in linear response to the linear susceptibility  via the relation
\begin{eqnarray}\label{Transmission_chi}
\mathcal T(\omega_{\rm p})=1-{\rm i}\mathcal{N}\hbar\omega_{\rm p}  \chi(\omega_{\rm p})\;.
\end{eqnarray}
\indent A this point we use the GME (\ref{GME}) to find the explicit expression for $p_1^{(1)}$ in terms of the NIBA kernels.
By substituting Eq.~(\ref{Pas}) and its time derivative in the limit $t\rightarrow \infty$ of the GME~\eqref{GME}, we arrive at the closed expression for $p_1^{(1)}$ ~\cite{Grifoni1995,Grifoni1998}
\begin{eqnarray}\label{Sp1SM}
p_1^{(1)}(\omega_{\rm p})&=&\frac{1}{{\rm i}\omega_{\rm p}+v^{+(0)}(\omega_{\rm p})}\left[ k^{-(1)}_1(\omega_{\rm p})-k^{+(1)}_{1}(\omega_{\rm p})\frac{k^{-(0)}_0}{k^{+(0)}_0}\right]\;,
\end{eqnarray}
where the superscripts $0$ and $1$ denote zeroth and first order in $\varepsilon_{\rm p}/\omega_{\rm p}$, respectively. Note that this expression for $p_1^{(1)}$ does not descend from a  Markovian limit of the GME, which  would yield $v^{+(0)}(0)$ instead of $v^{+(0)}(\omega_{\rm p})$ at the denominator in the prefactor. The kernels $k_m^{\pm}$ and $v^+$ in Eq.~(\ref{Sp1SM}) read
\begin{equation}\begin{aligned}\label{kv}
k_m^{\pm}(\omega_{\rm p})=&\frac{\omega_{\rm p}}{2\pi}\int_{-\pi/\omega_{\rm p}}^{\pi/\omega_{\rm p}}dt\;e^{-{\rm i}m\omega_{\rm p} t}\int_0^{\infty} d\tau\; \mathcal{K}^{\pm}(t,t-\tau)\;,\\
v^{+}(\omega_{\rm p})=&\frac{\omega_{\rm p}}{2\pi}\int_{-\pi/\omega_{\rm p}}^{\pi/\omega_{\rm p}}dt\;\int_0^{\infty} d\tau\; e^{-{\rm i}\omega_{\rm p} \tau}\mathcal{K}^{\pm}(t,t-\tau)\;,\\
\end{aligned}\end{equation}
where $\mathcal{K}^{\pm}(t,t')$ are defined in Eq.~(\ref{K}). Due to the integrations over the probe period $2\pi/\omega_{\rm p}$, the dynamical phase $\zeta(t,t')$  in $\mathcal{K}^\pm$ (see  Eq. (\ref{zeta})) yields the Bessel functions $J_m[(2\varepsilon_{\rm p}/\omega_{\rm p})\sin(\omega_{\rm p}t/2)]$ in the coefficients of  $k_m^{\pm}$ and $v^+$. The small amplitude of the probe field allows for expanding to lowest order in $\varepsilon_{\rm p}/\omega_{\rm p}$ the Bessel function by means of  $J_m(x)\sim (x/2)^m$, obtaining the following explicit expressions 
\begin{equation}\begin{aligned}\label{kpmv}
k^{+(0)}_{0}=&\int_0^{\infty}dt\; h^{+}(t)\cos(\varepsilon_0 t)\;,\\
k^{-(0)}_{0}=&\int_0^{\infty}dt\;h^{-}(t)\sin(\varepsilon_0 t)\;, \\
k^{+(1)}_{1}(\omega_{\rm p})=&-\frac{\varepsilon_{\rm p}}{\omega_{\rm p}}\int_0^{\infty}dt\;e^{-{\rm i}\omega_{\rm p} t/2}h^{+}(t)\sin(\varepsilon_0 t)\sin(\omega_{\rm p} t/2)\;,\\
k^{-(1)}_{1}(\omega_{\rm p})=&\frac{\varepsilon_{\rm p}}{\omega_{\rm p}}\int_0^{\infty}dt\;e^{-{\rm i}\omega_{\rm p} t/2}h^{-}(t)\cos(\varepsilon_0 t)\sin(\omega_{\rm p} t/2)\;,\\
{\rm and}\quad v^{+(0)}(\omega_{\rm p})=&\int_0^{\infty}dt\;e^{-{\rm i}\omega_{\rm p} t}h^{+}(t)\cos(\varepsilon_0 t)\;.
\end{aligned}\end{equation}
Within the present linear response treatment, the transmission is independent of the probe amplitude $\varepsilon_{\rm p}$, see Eq.~(\ref{Transmission_chi}). Note that, while the theory presented here assumes a small probe amplitude, it allows to describe  the situation in which the qubit is  strongly coupled to its environment. Moreover, the transmission spectrum of the qubit can be calculated also in the presence of a pump drive, as in \citep{Magazzu2018}, at least in the regime where the drive frequency is much larger than the renormalized value $\Delta_r$ of the qubit parameter $\Delta$. This condition is not restrictive in the strong coupling regime to Ohmic or sub-Ohmic baths ($G(\omega)\propto\omega^s$, with $s\leq 1$), which yields a strong renormalization and thus a small $\Delta_r$~\cite{Weiss2012}. At finite temperatures, the strong coupling regime makes the NIBA perform satisfactorily also in the presence of a static bias.

\section{Results}
\label{results}
In this section we apply the formalism reviewed above to the setup shown in Fig.~\ref{scheme}-a, where the static bias is zero, $\varepsilon_0=0$. This entails that $k^{-(0)}_{0}=k^{+(1)}_{1}=0$ in Eq. (\ref{kpmv}). The resulting expression for the susceptibility simplifies to 
\begin{eqnarray}\label{chi}
\chi(\omega_{\rm p})&=&\frac{1}{\hbar\varepsilon_{\rm p}}\frac{k^{-(1)}_1(\omega_{\rm p})}{{\rm i}\omega_{\rm p}+v^{+(0)}(\omega_{\rm p})}\;.
\end{eqnarray}
The effective spectral density in Eq. (\ref{Geff}), yields for the real and imaginary parts $Q'(t)$ and $Q''(t)$ of the bath correlation function in Eq. (\ref{Qdef}) the exlpicit expressions~\cite{Nesi2007} 
\begin{eqnarray}
\label{Q}
Q'(t)&=& X \tau - L \left(e^{-\Gamma t}\cos{\bar{\Omega}\tau}-1\right) - Z e^{-\Gamma t}\sin{\bar{\Omega}t}+Q'_{\rm Mats}(t)\;, \\
Q''(t)&=& \pi\alpha-e^{-\Gamma t} \pi\alpha \left(\cos{\bar{\Omega}t}-N\sin{\bar{\Omega}t}\right),
\end{eqnarray}
with $X=2\pi \alpha k_B T/\hbar$ and $\bar{\Omega}=\sqrt{\Omega^2-\Gamma^2}$ and where
\begin{equation}\begin{aligned}\label{}
N&=\frac{\Omega^2-2\Gamma^2}{2\Gamma\bar\Omega}\;,\\
L&=\pi \alpha\frac{N\sinh{(\beta\hbar\bar{\Omega})}-\sin{\left(\beta\hbar\Gamma\right)}}{\cosh{(\beta\hbar\bar{\Omega})}-\cos{\left(\beta\hbar\Gamma\right)}}\;,  \\
Z&=\pi \alpha\frac{\sinh{(\beta\hbar\bar{\Omega})}+N\sin{\left(\beta\hbar\Gamma\right)}}{\cosh{(\beta\hbar\bar{\Omega})}-\cos{\left(\beta\hbar\Gamma\right)}}\;.
\end{aligned}\end{equation}
The term $Q'_{\rm Mats}(t)$ is the following series over the Matsubara frequencies $\nu_n := n\;2\pi k_B T/\hbar$
\begin{equation}\label{Qmats}
Q'_{\rm Mats}(t)=4 \pi\alpha \frac{\Omega^4}{\hbar\beta}\sum_{n=1}^{+\infty} \dfrac{1}{(\Omega^2+\nu_n^2)^2-4\Gamma^2 \nu_n^2} \left[\dfrac{1-e^{-\nu_n t}}{\nu_n}\right]\;.
\end{equation}
\begin{figure}[h!]
\begin{center}
\hspace{-1.75cm}
\includegraphics[width=0.7\textwidth,angle=0]{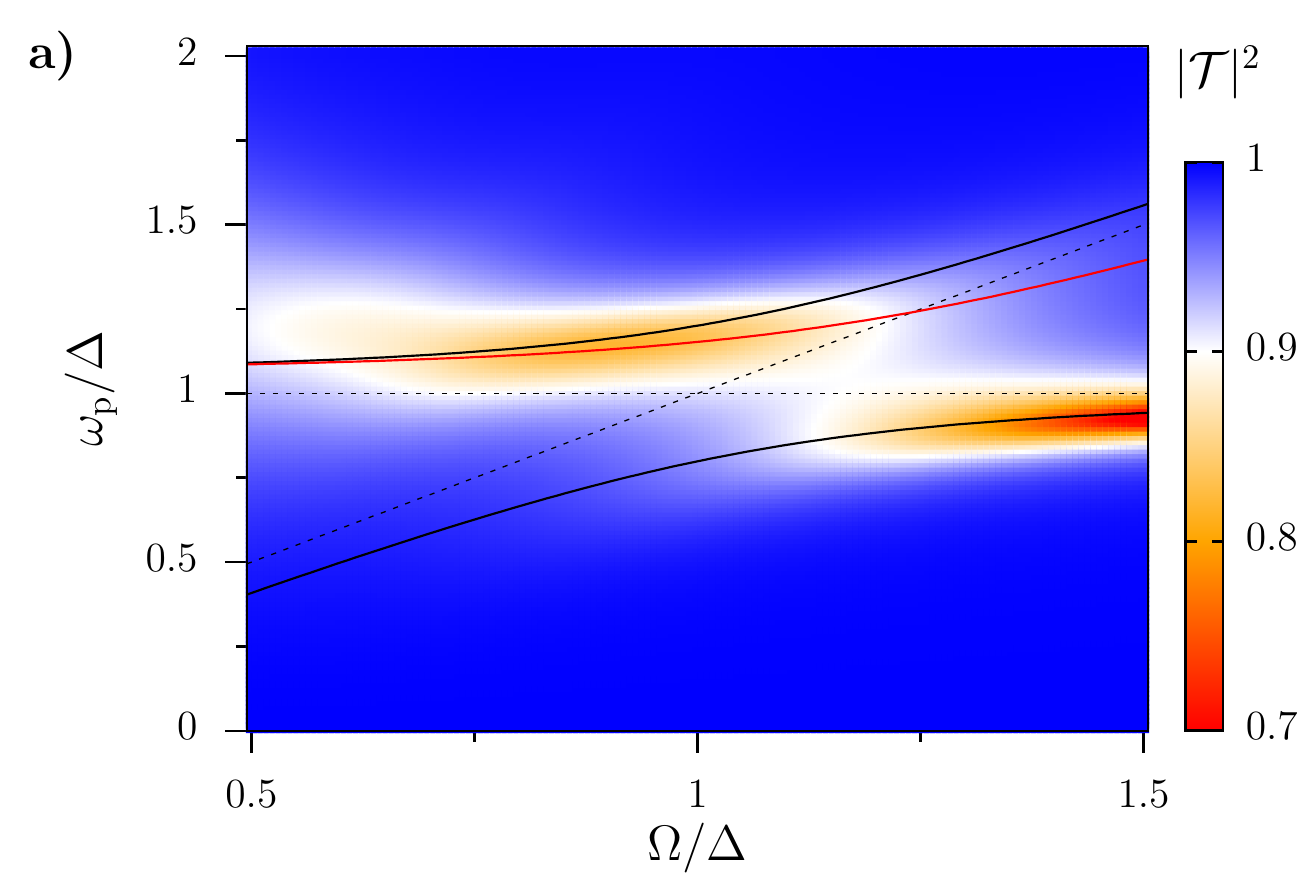}
\hspace{-0.5cm}
\includegraphics[width=0.4\textwidth,angle=0]{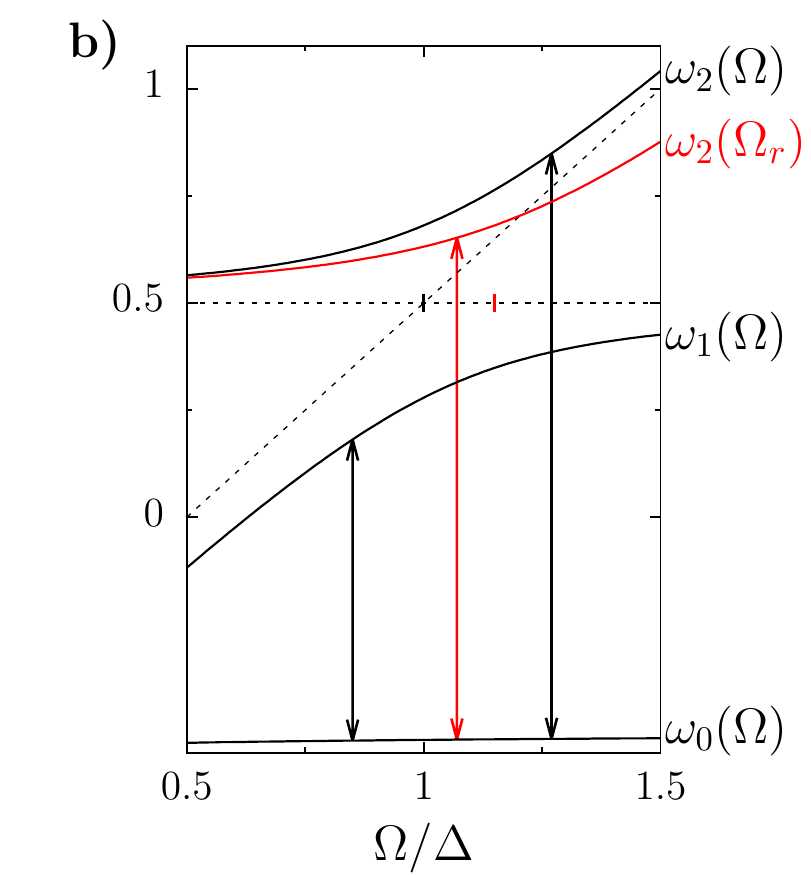}
\hspace{-1.5cm}
\caption{\small{\textbf{a} - Transmission $|\mathcal{T}|^2$ as a function of the resonator frequency $\Omega$ and probe frequency $\omega_{\rm p}$. Parameters are $g=0.2~\Delta$, $\kappa=0.05$, $T=\hbar\Delta/k_{\rm B}$, and $\mathcal{N}=0.1$. Black solid  lines - Transition frequencies $\omega_{10}(\Omega)$ and $\omega_{20}(\Omega)$, from the spectrum of the non-dissipative Rabi model, see Eq.~\eqref{Rabi_spectrum}.  Red solid line - Transition frequency  $\omega_{20}(\Omega_r)$, where we assumed  $\Omega_r=0.87~\Omega$, which yields the resonance condition at $\Omega\simeq 1.15~\Delta$. Dashed lines - transition frequencies for the decoupled qubit-resonator system, $g=0$. \textbf{b} - Eigenfrequencies of the coupled and uncoupled system and transition frequencies (arrows). The black and red tics on the horizontal dashed line indicate the resonance conditions $\Omega=\Delta$ and $\Omega_r=\Delta$, respectively.
Note that Eq.~\eqref{Rabi_spectrum} is a perturbative result valid for $g\ll \Delta, \Omega$.}}
\label{colormap}
\end{center}
\end{figure}
In what follows, we set the parameter that relates, according to  Eq. (\ref{Transmission_chi}), the calculated susceptibility to the measured transmission $\mathcal{T}$ at the probe frequency to the value $\mathcal{N}=0.1$. Moreover, while varying the resonator frequency $\Omega$ and qubit-resonator coupling $g$, we fix the resonator-bath coupling  to $\kappa=0.05$. Finally, the temperature of the bath is chosen to be $T=\hbar\Delta/k_B$.\\
\indent In Fig.~\ref{colormap}, we show the full qubit transmission spectrum with the qubit-resonator coupling set to $g=0.2~\Delta$, namely in the USC regime. Specifically, the transmission $|\mathcal{T}|^2$ is calculated as a function of the oscillator frequency $\Omega$ and of the probe frequency $\omega_{\rm p}$. 
The transition frequencies $\omega_{10}=\omega_{1}-\omega_{0}$ and $\omega_{20}=\omega_{2}-\omega_{0}$ of the non-dissipative model, from Eq.~\eqref{Rabi_spectrum}, are also shown along with the corresponding quantities for the uncoupled system, $g=0$, to highlight the presence of the avoided crossing at the resonance condition $\Omega=\Delta$. In the regions where the transmission is not complete, $|\mathcal{T}|^2<1$, the qubit response to the probe is different from zero, meaning that the qubit dynamics has a component at the probe frequency.\\ 
\indent To appreciate the features of the transmission plotted in Fig.~\ref{colormap}, we first describe the features of the qubit dynamics given by the population difference in Fourier space, $F(\omega)=2\int_0^\infty dt\; \cos(\omega t)P(t)$,  as analyzed in~\cite{Hausinger2008} for the non-dissipative and the weakly dissipative cases in the absence of a probe field. In the non-dissipative case, $F(\omega)$ is characterized by a sequence of peaks with two dominating contributions at $\omega_{10}$ and $\omega_{20}$, the two frequencies being separated, at resonance, by the vacuum Rabi splitting $2g$. In the presence of a weak dissipation, $\kappa=0.015$ and $T=0.1~\hbar\Delta/k_{\rm B}$ , the secondary peaks are washed out and the two main peaks are broadened. Moreover, the relative magnitude of the peaks depends on the detuning $\Delta-\Omega$. This can be accounted for within a Bloch-Redfield master equation approach: One finds that the contribution to $F(\omega)$ at  frequency $\omega_{n0}$, with $n=1,2$, is weighted by the factor $\Gamma_{n0}^{-1}$, where $\Gamma_{nm}$ are the dephasing rates in the full secular approximation. Evaluation of the dephasing rates shows that negative detuning  yields $\Gamma_{10}>\Gamma_{20}$ and thus a larger contribution of the peak at the higher frequency, whereas positive detuning yields a dominating peak at the lower frequency.\\   
\indent These features  are qualitatively present in the corresponding behavior of the transmission shown in Fig.~\ref{colormap} for the stronger dissipation/higher temperature regime considered here. There are however interesting peculiarities that arise from the interplay between the detuning and dissipation. 
Indeed, due to the coupling to the heat bath,  the resonator frequency is renormalized to $\Omega_r<\Omega$. As a result, the resonance condition $\Omega_r=\Delta$ occurs at some value of $\Omega$ larger than $\Delta$. This is reflected by the fact that the simultaneous presence of two dips in the transmission, expected at $\Omega\simeq\Delta$ for weak dissipation, here occurs around the value $1.15~\Delta$. The renormalization of the oscillator frequency also accounts for the fact that, for $\Omega\gtrsim\Delta$, the trace of the dip at the lower frequency is well reproduced by the curve $\omega_{10}(\Omega)$ while the one at the higher frequency is reproduced by $\omega_{20}(\Omega_r)$, where we assume the simple relation $\Omega_r=0.87~\Omega$ in order to have the resonant condition at $\Omega\simeq 1.15~\Delta$, see the red solid line in Fig.~\ref{colormap}. The  shift of the dip positions is non-symmetric because the transition frequency  $\omega_{20}$ is more affected by the renormalization of the oscillator frequency than $\omega_{10}$. The reason is  that the dominant contributions to the eigenfrequencies in 
Eq.~\eqref{Rabi_spectrum} come from the uncoupled case, $g=0$, which gives $\omega_{20}=\Omega$ and $\omega_{10}=\Delta$ at positive detuning ($\Omega>\Delta$). As a result of this non-symmetric shift, around the resonance the distance between the dips is less than the vacuum Rabi splitting $2g$.
The renormalization towards lower values of the oscillator frequency also enhances the loss of accuracy at low $\Omega$ of the perturbative calculation ($g\ll \Delta, \Omega$) yielding the eigenfrequencies in Eq.~\eqref{Rabi_spectrum}.    
\begin{figure}[ht!]
\begin{center}
\hspace{-0.5cm}
\includegraphics[width=0.525\linewidth,angle=0]{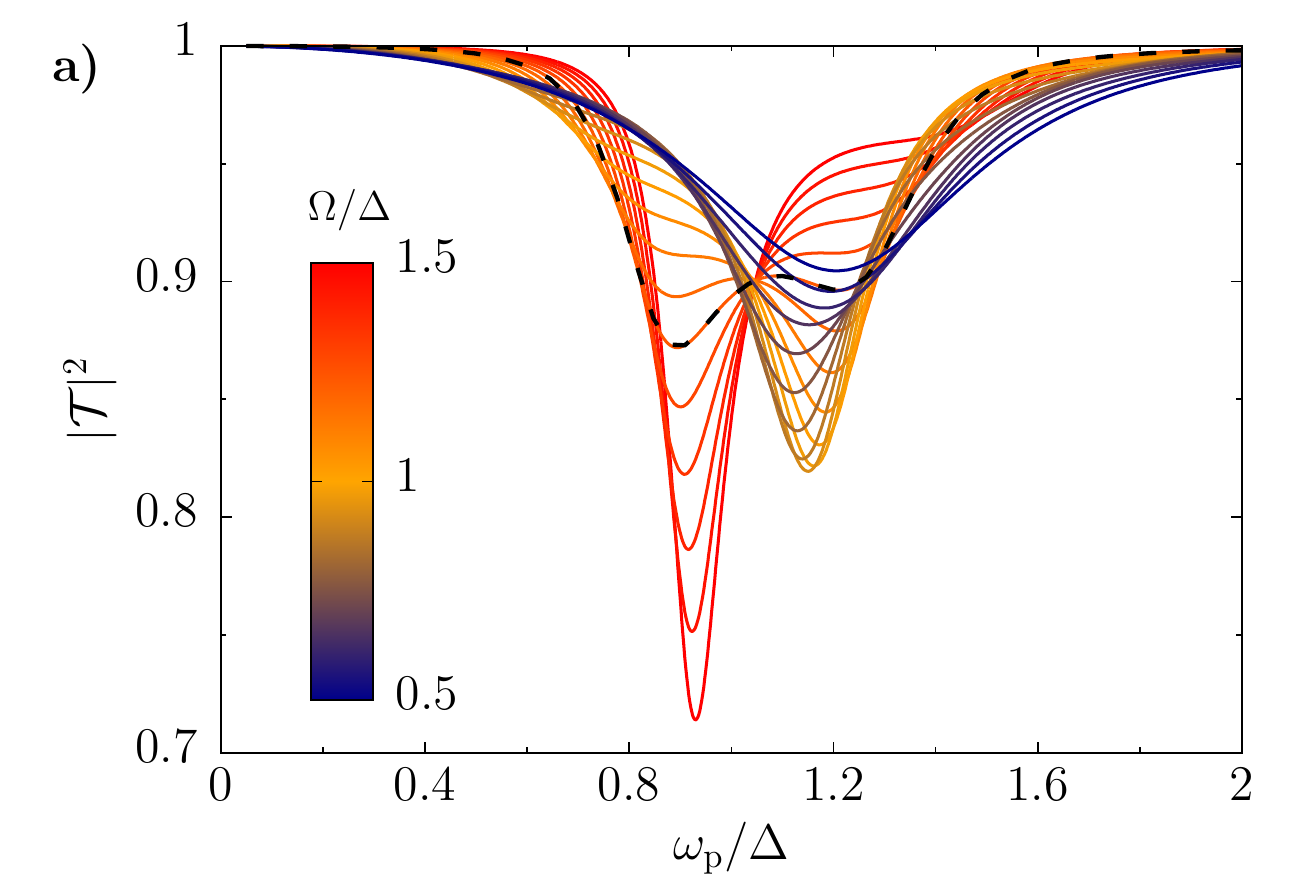}
\hspace{-0.7cm}
\includegraphics[width=0.525\linewidth,angle=0]{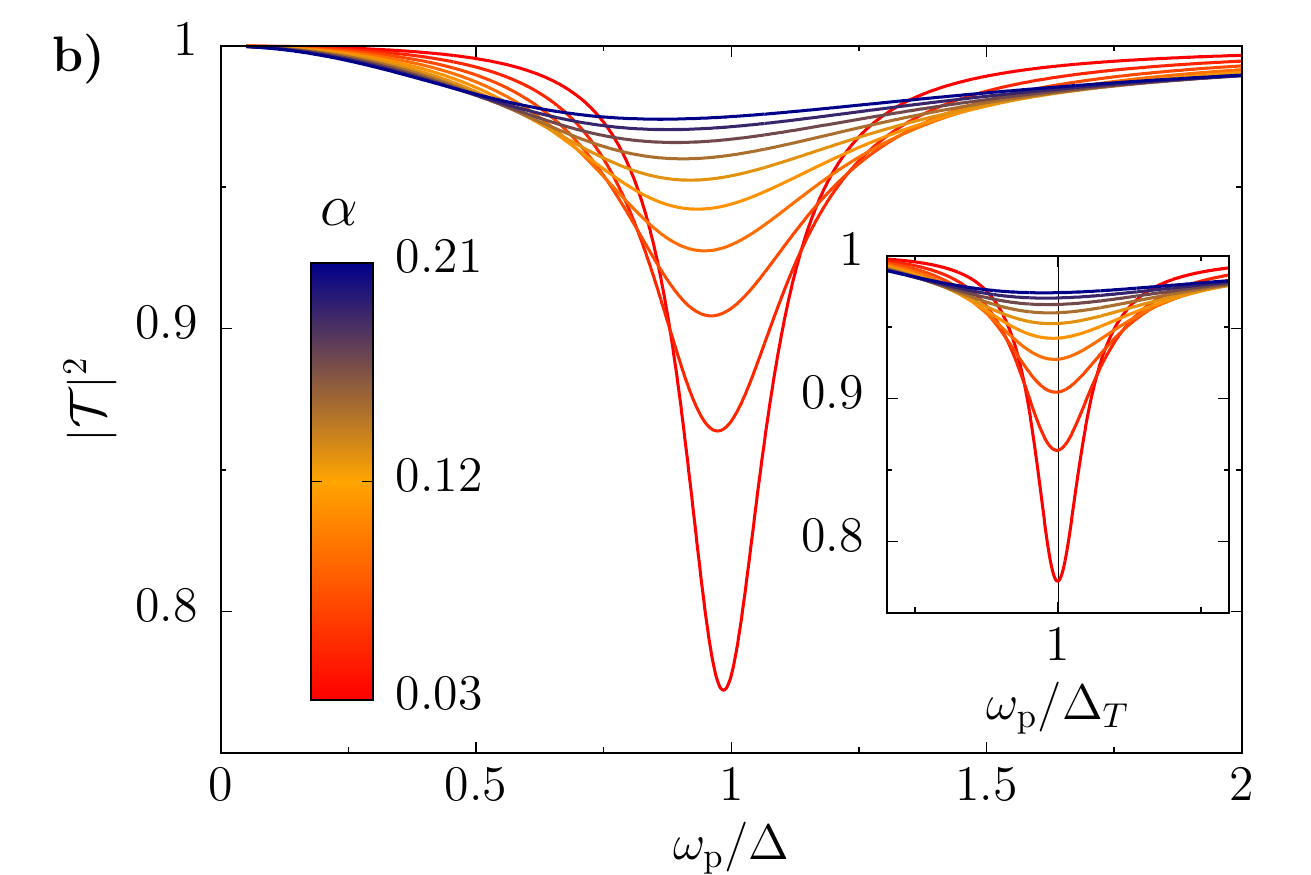}
\caption{\small{Transmission $|\mathcal{T}|^2$ as a function of the probe frequency $\omega_{\rm p}$. \textbf{a} - Qubit-dissipative resonator system for different values of the resonator frequency $\Omega$. Parameters are the same as in Fig.~\ref{colormap}. The relative weight of the dips is dictated by the detuning with respect to the renormalized resonator frequency $\Omega_r$. The dip at high (low) frequency dominates for $\Omega_r<\Delta$ ( $\Omega_r > \Delta$).  The dashed line is at $\Omega=1.2~\Delta$.
\textbf{b} - Qubit directly coupled to an Ohmic bath for different values of the coupling strength $\alpha$. Bath temperature and cutoff frequency are $T=\hbar\Delta/k_B$ and $\omega_{\rm c}=10\;\Delta$, respectively, and $\mathcal{N}=0.1$. Inset - Transmission \emph{vs}. the probe frequency scaled with the temperature-dependent renormalized qubit splitting $\Delta_T$ of Eq.~\eqref{Delta_T}: The dips in the transmission are centered at $\omega_{\rm p}=\Delta_T$.}}
\label{curves}
\end{center}
\end{figure}
 At the extrema of the $\Omega$-range we notice that just one of the two dips in the transmission survives. At small resonator frequencies, $\Omega\simeq0.5~\Delta$, the transmission displays a single, broad dip centered at $\omega_{\rm p} \simeq 1.2~\Delta$. The large broadening is consistent with the fact that, by decreasing  $\Omega$, the effective coupling $\alpha$ between the qubit and the structured spectral density in Eq.~\eqref{Geff} increases. In the opposite limit,
around $\Omega=1.5~\Delta$, there is a single, narrow dip centered towards $\Omega=\Delta$, consistently with the fact that at large $\Omega$ the effective spectral density of Eq.~\eqref{Geff} reproduce a weakly coupled Ohmic bath. \\
\indent These features are best seen in Fig.~\ref{curves}-a, where the transmission is shown as a function of the probe frequency for different values of the resonator frequency $\Omega$, from negative to positive detuning $\Omega_r-\Delta$, with the same parameters as in the colormap of Fig.~\ref{colormap}. The curves show, around the resonance condition  $\Omega_r \simeq \Delta$, the (broadened) two-dip pattern characteristic of the qubit-oscillator system.  At large values of $\Omega$ the spectra  present a single narrow dip at a frequency which tends to the position $\omega_{\rm p}=\Delta$, as the effective coupling $\alpha$ decreases by increasing $\Omega$. In the opposite regime of small $\Omega$, the effective coupling $\alpha$ tends to be large and the doubly peaked structure is smoothed out to leave a single broad dip centered at $\omega_{\rm p}\simeq 1.2~\Delta$. This \emph{upwards} renormalization of the qubit splitting due to the structured environment beyond weak dissipation has been already observed in~\cite{vonDelft2004} using the flow-equation renormalization group approach. It is in striking contrast with the downwards renormalization of $\Delta$ occurring in the Ohmic case, whereby upon increasing the coupling to the heat bath, the dip in the transmission also broadens but tends to lower frequencies. This point is exemplified by Fig.~\ref{curves}-b where, for comparison, the transmission is shown for the qubit directly coupled to a Ohmic bath with spectral density function $G_{\rm Ohmic}=2\alpha\omega\exp(-\omega/\omega_{\rm c})$, for different values of the qubit-bath coupling $\alpha$.
\begin{figure}[h!]
\begin{center}
\includegraphics[width=0.65\textwidth,angle=0]{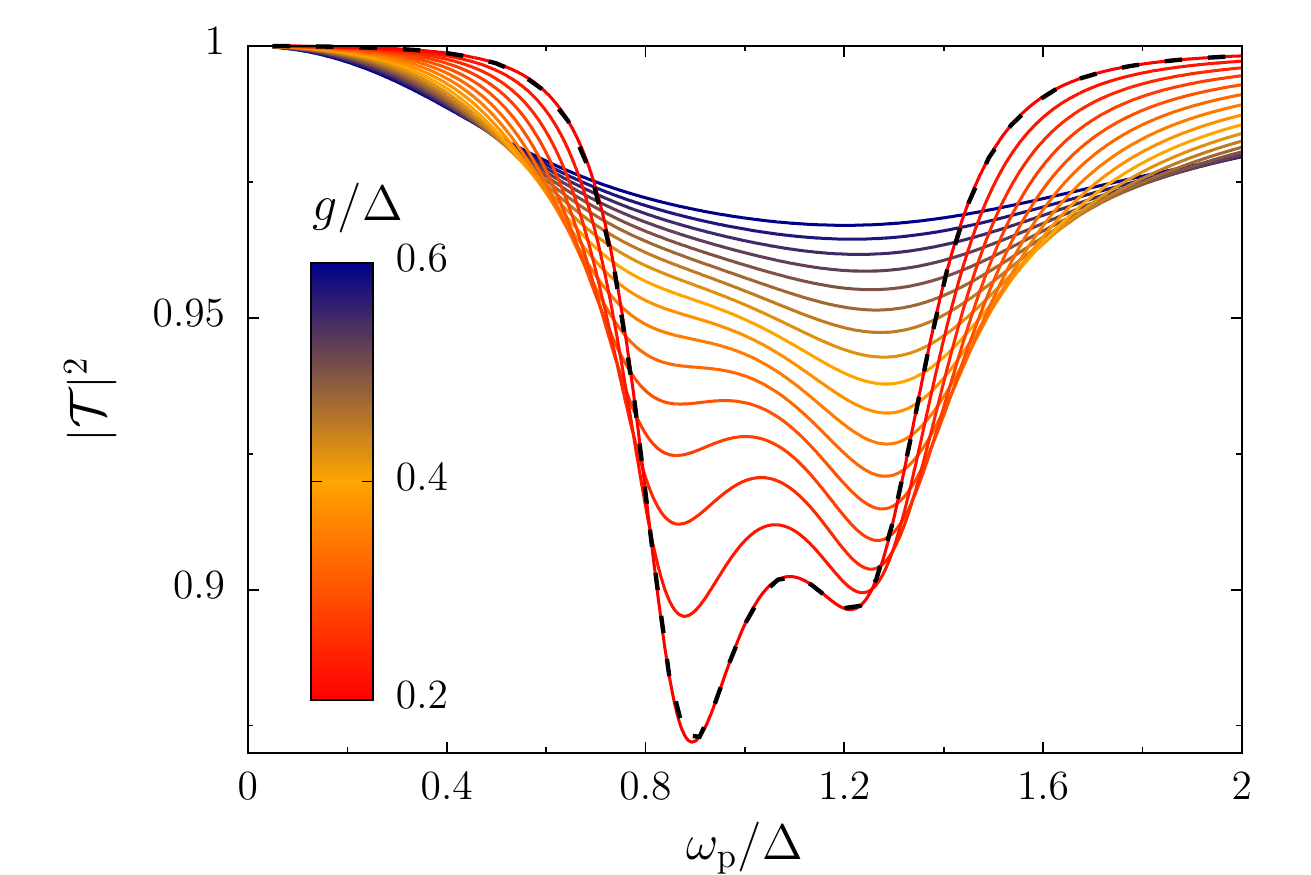}
\caption{\small{Transmission $|\mathcal{T}|^2$ as a function of the probe frequency $\omega_{\rm p}$ for different values of the qubit-resonator coupling $g$. The resonators frequency is $\Omega=1.2\;\Delta$ and the remaining parameters are as in Fig.~\ref{colormap}. The separation between the peaks increases as $g$ is increased.  The dashed curves here and in Fig.~\ref{curves}-a are obtained with the same set of parameters and thus coincide. Increasing $g$, the separation between the dips increases and the dip at high frequency becomes the dominant.}}
\label{curves_g}
\end{center}
\end{figure}
In the temperature/coupling regime considered in Fig.~\ref{curves}-b, the temperature-dependent renormalization of the qubit frequency splitting for the Ohmic bath is given by~\cite{Weiss2012}
 \begin{equation}
\label{Delta_T}
\Delta_T=\Delta_{\rm r}(\nu_1/\Delta_{\rm r})^\alpha\;,
\end{equation}
with $\Delta_{\rm r}=\Delta(\Delta/\omega_{\rm c})^{\alpha/(1-\alpha)}$ the renormalized splitting at $T=0$ and $\nu_1=2\pi k_{\rm B}T/\hbar$ the first Matsubara frequency. As shown in the inset of Fig.~\ref{curves}-b, by scaling the probe frequency with this renormalized splitting, the positions of the dips at different $\alpha$ collapse to the value 1.\\
\indent We complement the information provided in Figs.~\ref{colormap} and \ref{curves} with the curves of Fig.~\ref{curves_g}, where the qubit-resonator coupling $g$ is increased from $0.2$ to $0.6$, in units of $\Delta$, with $\Omega$ fixed to the value $1.2\;\Delta$. While for the lower values of  coupling the dip the low-frequency dip is dominant, by increasing $g$, the dip at the higher frequency dominates. At large qubit-resonator coupling, the two-dip structure is smoothed out resulting in a single, broad dip centered at $\omega_{\rm p}>\Delta$, similarly to what happens by decreasing $\Omega$ at fixed $g$.\\   
\indent The present analysis can be pushed towards lower coupling strengths and temperatures, where the NIBA is still reliable for the unbiased system, see for example the analytical weak damping approximation, derived within NIBA in~\cite{Nesi2007}. Moreover, at low temperatures and weak qubit-oscillator coupling, an analytical treatment beyond the rotating-wave approximation, which accounts also for a static bias, $\varepsilon_0\neq 0$, is provided in~\cite{Hausinger2008}.\\   
\indent As a final remark, we note that, in the present linear response regime to the probe field, the weak time-dependent bias does not spoil the NIBA results. This is because linear response theory reproduces the dissipative dynamics of the static system which is in turn well-described within the NIBA. In~\cite{Goorden2004}, the performance of the NIBA with an effective spectral density of the same type as Eq.~\eqref{Geff} has been compared with the numerically exact results of QUAPI.

\section{Conclusions}
In recent years, superconducting quantum circuits assumed the role of leading platforms for quantum computing and simulations. In the latter context, a qubit coupled to a superconducting resonator forms the basic setup for quantum optics experiments in the ultrastrong coupling regime of light and matter. To account for the presence of dissipation, brought in by the coupling of the resonator to  a reservoir of bosonic modes, the system can be mapped to a spin-boson system with the distribution of environmental couplings displaying a peak at the oscillator frequency, rendering a so-called structured bath.
The noninteracting-blip approximation represents a valuable tool for investigating the qubit reduced dynamics in the presence of dissipation and in the nonperturbative regime of qubit-resonator coupling.\\
\indent In our work, we employed the tools developed within this approximation scheme to spectroscopically analyze the setup of a flux qubit ultrastrongly coupled to a dissipative resonator. The transmission of a weak probe field, which is measured in actual experiments, is connected to the qubit dynamics under the influence of the resonator-bath system, via a linear response treatment.    
We investigated how dissipation in the resonator affects the qubit transmission spectra by varying the qubit-resonator detuning and its coupling to the qubit. The interaction between the resonator and a heat bath beyond the weak coupling limit introduces a renormalization of the resonator frequency which, in turn, modifies the qubit response. We find a bath-induced shift of the resonance condition and a decreased vacuum Rabi splitting, due to the fact that the renormalization of the oscillator frequency affects the relevant transitions in a non-symmetric way. Moreover, the broadening of the dips in the transmission, which is due to dissipation, depends on the qubit-resonator detuning. Finally, an upwards renormalization effect of the qubit splitting is found at small resonator frequencies, witnessing the structured nature of the effective environment of the qubit. \\ 
\indent The study performed here can be extended to multiple baths and different coupling setups and the formalism can account for the presence of driving in the qubit parameters. In recent years, a two-bath version of the spin-boson model has been explored in the context of  heat transfer between heat baths~\cite{Segal2011,Segal2014, Segal2014PRE, Ankerhold2015, Carrega2015,Carrega2016, Wang2017,Agarwalla2017,Nazir2017}. The formalism used here is suitable for describing a setup for heat transport in the quantum regime of the same kind of the one realized in~\cite{Pekola2018}, where a qubit is connected to two heat baths at different temperatures via waveguide resonators.\\  
\section*{Acknowledgments}
The authors acknowledge financial
support by the Deutsche Forschungsgemeinschaft via  SFB 1277 B02.


\begin{thebibliography}{49}%
\makeatletter
\providecommand \@ifxundefined [1]{%
 \@ifx{#1\undefined}
}%
\providecommand \@ifnum [1]{%
 \ifnum #1\expandafter \@firstoftwo
 \else \expandafter \@secondoftwo
 \fi
}%
\providecommand \@ifx [1]{%
 \ifx #1\expandafter \@firstoftwo
 \else \expandafter \@secondoftwo
 \fi
}%
\providecommand \natexlab [1]{#1}%
\providecommand \enquote  [1]{``#1''}%
\providecommand \bibnamefont  [1]{#1}%
\providecommand \bibfnamefont [1]{#1}%
\providecommand \citenamefont [1]{#1}%
\providecommand \href@noop [0]{\@secondoftwo}%
\providecommand \href [0]{\begingroup \@sanitize@url \@href}%
\providecommand \@href[1]{\@@startlink{#1}\@@href}%
\providecommand \@@href[1]{\endgroup#1\@@endlink}%
\providecommand \@sanitize@url [0]{\catcode `\\12\catcode `\$12\catcode
  `\&12\catcode `\#12\catcode `\^12\catcode `\_12\catcode `\%12\relax}%
\providecommand \@@startlink[1]{}%
\providecommand \@@endlink[0]{}%
\providecommand \url  [0]{\begingroup\@sanitize@url \@url }%
\providecommand \@url [1]{\endgroup\@href {#1}{\urlprefix }}%
\providecommand \urlprefix  [0]{URL }%
\providecommand \Eprint [0]{\href }%
\providecommand \doibase [0]{http://dx.doi.org/}%
\providecommand \selectlanguage [0]{\@gobble}%
\providecommand \bibinfo  [0]{\@secondoftwo}%
\providecommand \bibfield  [0]{\@secondoftwo}%
\providecommand \translation [1]{[#1]}%
\providecommand \BibitemOpen [0]{}%
\providecommand \bibitemStop [0]{}%
\providecommand \bibitemNoStop [0]{.\EOS\space}%
\providecommand \EOS [0]{\spacefactor3000\relax}%
\providecommand \BibitemShut  [1]{\csname bibitem#1\endcsname}%
\let\auto@bib@innerbib\@empty
\bibitem [{\citenamefont {You}\ and\ \citenamefont {Nori}(2011)}]{You2011}%
  \BibitemOpen
  \bibfield  {author} {\bibinfo {author} {\bibfnamefont {J.~Q.}\ \bibnamefont
  {You}}\ and\ \bibinfo {author} {\bibfnamefont {F.}~\bibnamefont {Nori}},\
  }\bibfield  {title} {\enquote {\bibinfo {title} {{Atomic physics and quantum
  optics using superconducting circuits}},}\ }\href {\doibase
  10.1038/nature10122} {\bibfield  {journal} {\bibinfo  {journal} {Nature}\
  }\textbf {\bibinfo {volume} {474}},\ \bibinfo {pages} {589--597} (\bibinfo
  {year} {2011})}\BibitemShut {NoStop}%
\bibitem [{\citenamefont {Houck}\ \emph {et~al.}(2012)\citenamefont {Houck},
  \citenamefont {T{\"u}reci},\ and\ \citenamefont {Koch}}]{Koch2012}%
  \BibitemOpen
  \bibfield  {author} {\bibinfo {author} {\bibfnamefont {A.~A.}\ \bibnamefont
  {Houck}}, \bibinfo {author} {\bibfnamefont {H.~E.}\ \bibnamefont
  {T{\"u}reci}}, \ and\ \bibinfo {author} {\bibfnamefont {J.}~\bibnamefont
  {Koch}},\ }\bibfield  {title} {\enquote {\bibinfo {title} {{On-chip quantum
  simulation with superconducting circuits}},}\ }\href@noop {} {\bibfield
  {journal} {\bibinfo  {journal} {Nat. Phys.}\ }\textbf {\bibinfo {volume}
  {8}},\ \bibinfo {pages} {292} (\bibinfo {year} {2012})}\BibitemShut {NoStop}%
\bibitem [{\citenamefont {Pekola}(2015)}]{Pekola2015}%
  \BibitemOpen
  \bibfield  {author} {\bibinfo {author} {\bibfnamefont {J.~P.}\ \bibnamefont
  {Pekola}},\ }\bibfield  {title} {\enquote {\bibinfo {title} {{Towards quantum
  thermodynamics in electronic circuits}},}\ }\href@noop {} {\bibfield
  {journal} {\bibinfo  {journal} {Nat. Phys.}\ }\textbf {\bibinfo {volume}
  {11}},\ \bibinfo {pages} {118} (\bibinfo {year} {2015})}\BibitemShut
  {NoStop}%
\bibitem [{\citenamefont {Wendin}(2017)}]{Wendin2017}%
  \BibitemOpen
  \bibfield  {author} {\bibinfo {author} {\bibfnamefont {G.}~\bibnamefont
  {Wendin}},\ }\bibfield  {title} {\enquote {\bibinfo {title} {{Quantum
  information processing with superconducting circuits: a review}},}\
  }\href@noop {} {\bibfield  {journal} {\bibinfo  {journal} {Rep. Prog. Phys.}\
  }\textbf {\bibinfo {volume} {80}},\ \bibinfo {pages} {106001} (\bibinfo
  {year} {2017})}\BibitemShut {NoStop}%
\bibitem [{\citenamefont {Gu}\ \emph {et~al.}(2017)\citenamefont {Gu},
  \citenamefont {Kockum}, \citenamefont {Miranowicz}, \citenamefont {Liu},\
  and\ \citenamefont {Nori}}]{Nori2017}%
  \BibitemOpen
  \bibfield  {author} {\bibinfo {author} {\bibfnamefont {X.}~\bibnamefont
  {Gu}}, \bibinfo {author} {\bibfnamefont {A.~F.}\ \bibnamefont {Kockum}},
  \bibinfo {author} {\bibfnamefont {A.}~\bibnamefont {Miranowicz}}, \bibinfo
  {author} {\bibfnamefont {Y.-X.}\ \bibnamefont {Liu}}, \ and\ \bibinfo
  {author} {\bibfnamefont {F.}~\bibnamefont {Nori}},\ }\bibfield  {title}
  {\enquote {\bibinfo {title} {{Microwave photonics with superconducting
  quantum circuits}},}\ }\href@noop {} {\bibfield  {journal} {\bibinfo
  {journal} {Phys. Rep.}\ }\textbf {\bibinfo {volume} {718-719}},\ \bibinfo
  {pages} {1--102} (\bibinfo {year} {2017})}\BibitemShut {NoStop}%
\bibitem [{\citenamefont {Niemczyk}\ \emph {et~al.}(2010)\citenamefont
  {Niemczyk}, \citenamefont {Deppe}, \citenamefont {Huebl}, \citenamefont
  {Menzel}, \citenamefont {Hocke}, \citenamefont {Schwarz}, \citenamefont
  {Garcia-Ripoll}, \citenamefont {Zueco}, \citenamefont {H{\"u}mmer},
  \citenamefont {Solano}, \citenamefont {Marx},\ and\ \citenamefont
  {Gross}}]{Niemczyk2010}%
  \BibitemOpen
  \bibfield  {author} {\bibinfo {author} {\bibfnamefont {T.}~\bibnamefont
  {Niemczyk}}, \bibinfo {author} {\bibfnamefont {F.}~\bibnamefont {Deppe}},
  \bibinfo {author} {\bibfnamefont {H.}~\bibnamefont {Huebl}}, \bibinfo
  {author} {\bibfnamefont {P.~E.}\ \bibnamefont {Menzel}}, \bibinfo {author}
  {\bibfnamefont {F.}~\bibnamefont {Hocke}}, \bibinfo {author} {\bibfnamefont
  {J.~M.}\ \bibnamefont {Schwarz}}, \bibinfo {author} {\bibfnamefont {J.~J.}\
  \bibnamefont {Garc{\'i}a-Ripoll}}, \bibinfo {author} {\bibfnamefont
  {D.}~\bibnamefont {Zueco}}, \bibinfo {author} {\bibfnamefont
  {T.}~\bibnamefont {H{\"u}mmer}}, \bibinfo {author} {\bibfnamefont
  {E.}~\bibnamefont {Solano}}, \bibinfo {author} {\bibfnamefont
  {A.}~\bibnamefont {Marx}}, \ and\ \bibinfo {author} {\bibfnamefont
  {R.}~\bibnamefont {Gross}},\ }\bibfield  {title} {\enquote {\bibinfo {title}
  {{Circuit quantum electrodynamics in the ultrastrong-coupling regime}},}\
  }\href@noop {} {\bibfield  {journal} {\bibinfo  {journal} {Nat. Phys.}\
  }\textbf {\bibinfo {volume} {6}},\ \bibinfo {pages} {772} (\bibinfo {year}
  {2010})}\BibitemShut {NoStop}%
\bibitem [{\citenamefont {Ashhab}\ and\ \citenamefont
  {Nori}(2010)}]{Ashhab2010}%
  \BibitemOpen
  \bibfield  {author} {\bibinfo {author} {\bibfnamefont {S.}~\bibnamefont
  {Ashhab}}\ and\ \bibinfo {author} {\bibfnamefont {F.}~\bibnamefont {Nori}},\
  }\bibfield  {title} {\enquote {\bibinfo {title} {{Qubit-oscillator systems in
  the ultrastrong-coupling regime and their potential for preparing
  nonclassical states}},}\ }\href@noop {} {\bibfield  {journal} {\bibinfo
  {journal} {Phys. Rev. A}\ }\textbf {\bibinfo {volume} {81}},\ \bibinfo
  {pages} {042311} (\bibinfo {year} {2010})}\BibitemShut {NoStop}%
\bibitem [{\citenamefont {Forn-D{\'i}az}\ \emph {et~al.}(2010)\citenamefont
  {Forn-D{\'i}az}, \citenamefont {Lisenfeld}, \citenamefont {Marcos},
  \citenamefont {Garc{\'i}a-Ripoll}, \citenamefont {Solano}, \citenamefont
  {Harmans},\ and\ \citenamefont {Mooij}}]{Forn-Diaz2010}%
  \BibitemOpen
  \bibfield  {author} {\bibinfo {author} {\bibfnamefont {P.}~\bibnamefont
  {Forn-D{\'i}az}}, \bibinfo {author} {\bibfnamefont {J.}~\bibnamefont
  {Lisenfeld}}, \bibinfo {author} {\bibfnamefont {D.}~\bibnamefont {Marcos}},
  \bibinfo {author} {\bibfnamefont {J.~J.}\ \bibnamefont {Garc{\'i}a-Ripoll}},
  \bibinfo {author} {\bibfnamefont {E.}~\bibnamefont {Solano}}, \bibinfo
  {author} {\bibfnamefont {C.~J. P.~M.}\ \bibnamefont {Harmans}}, \ and\
  \bibinfo {author} {\bibfnamefont {J.~E.}\ \bibnamefont {Mooij}},\ }\bibfield
  {title} {\enquote {\bibinfo {title} {{Observation of the Bloch-Siegert Shift
  in a Qubit-Oscillator System in the Ultrastrong Coupling Regime}},}\
  }\href@noop {} {\bibfield  {journal} {\bibinfo  {journal} {Phys. Rev. Lett.}\
  }\textbf {\bibinfo {volume} {105}},\ \bibinfo {pages} {237001} (\bibinfo
  {year} {2010})}\BibitemShut {NoStop}%
\bibitem [{\citenamefont {Yoshihara}\ \emph {et~al.}(2016)\citenamefont
  {Yoshihara}, \citenamefont {Fuse}, \citenamefont {Ashhab}, \citenamefont
  {Kakuyanagi}, \citenamefont {Saito},\ and\ \citenamefont
  {Semba}}]{Yoshihara2017}%
  \BibitemOpen
  \bibfield  {author} {\bibinfo {author} {\bibfnamefont {F.}~\bibnamefont
  {Yoshihara}}, \bibinfo {author} {\bibfnamefont {T.}~\bibnamefont {Fuse}},
  \bibinfo {author} {\bibfnamefont {S.}~\bibnamefont {Ashhab}}, \bibinfo
  {author} {\bibfnamefont {K.}~\bibnamefont {Kakuyanagi}}, \bibinfo {author}
  {\bibfnamefont {S.}~\bibnamefont {Saito}}, \ and\ \bibinfo {author}
  {\bibfnamefont {K.}~\bibnamefont {Semba}},\ }\bibfield  {title} {\enquote
  {\bibinfo {title} {{Superconducting qubit-oscillator circuit beyond the
  ultrastrong-coupling regime}},}\ }\href@noop {} {\bibfield  {journal}
  {\bibinfo  {journal} {Nat. Phys.}\ }\textbf {\bibinfo {volume} {13}},\
  \bibinfo {pages} {44} (\bibinfo {year} {2016})}\BibitemShut {NoStop}%
\bibitem [{\citenamefont {Yoshihara}\ \emph {et~al.}(2017)\citenamefont
  {Yoshihara}, \citenamefont {Fuse}, \citenamefont {Ashhab}, \citenamefont
  {Kakuyanagi}, \citenamefont {Saito},\ and\ \citenamefont
  {Semba}}]{Yoshihara2017PRA}%
  \BibitemOpen
  \bibfield  {author} {\bibinfo {author} {\bibfnamefont {F.}~\bibnamefont
  {Yoshihara}}, \bibinfo {author} {\bibfnamefont {T.}~\bibnamefont {Fuse}},
  \bibinfo {author} {\bibfnamefont {S.}~\bibnamefont {Ashhab}}, \bibinfo
  {author} {\bibfnamefont {K.}~\bibnamefont {Kakuyanagi}}, \bibinfo {author}
  {\bibfnamefont {S.}~\bibnamefont {Saito}}, \ and\ \bibinfo {author}
  {\bibfnamefont {K.}~\bibnamefont {Semba}},\ }\bibfield  {title} {\enquote
  {\bibinfo {title} {Characteristic spectra of circuit quantum electrodynamics
  systems from the ultrastrong- to the deep-strong-coupling regime},}\ }\href
  {\doibase 10.1103/PhysRevA.95.053824} {\bibfield  {journal} {\bibinfo
  {journal} {Phys. Rev. A}\ }\textbf {\bibinfo {volume} {95}},\ \bibinfo
  {pages} {053824} (\bibinfo {year} {2017})}\BibitemShut {NoStop}%
\bibitem [{\citenamefont {Forn-D{\'i}az}\ \emph {et~al.}(2019)\citenamefont
  {Forn-D{\'i}az}, \citenamefont {Lamata}, \citenamefont {Rico}, \citenamefont
  {Kono},\ and\ \citenamefont {Solano}}]{Forn-Diaz2018review}%
  \BibitemOpen
  \bibfield  {author} {\bibinfo {author} {\bibfnamefont {P.}~\bibnamefont
  {Forn-D{\'i}az}}, \bibinfo {author} {\bibfnamefont {L.}~\bibnamefont
  {Lamata}}, \bibinfo {author} {\bibfnamefont {E.}~\bibnamefont {Rico}},
  \bibinfo {author} {\bibfnamefont {J.}~\bibnamefont {Kono}}, \ and\ \bibinfo
  {author} {\bibfnamefont {E.}~\bibnamefont {Solano}},\ }\bibfield  {title}
  {\enquote {\bibinfo {title} {{Ultrastrong coupling regimes of light-matter
  interaction}},}\ }\href
  {https://link.aps.org/doi/10.1103/RevModPhys.91.025005} {\bibfield  {journal}
  {\bibinfo  {journal} {Rev. Mod. Phys.}\ }\textbf {\bibinfo {volume} {91}},\
  \bibinfo {pages} {025005} (\bibinfo {year} {2019})}\BibitemShut {NoStop}%
\bibitem [{\citenamefont {Kockum}\ \emph {et~al.}(2019)\citenamefont {Kockum},
  \citenamefont {Miranowicz}, \citenamefont {{De Liberato}}, \citenamefont
  {Savasta},\ and\ \citenamefont {Nori}}]{Kockum2019}%
  \BibitemOpen
  \bibfield  {author} {\bibinfo {author} {\bibfnamefont {F.~A.}\ \bibnamefont
  {Kockum}}, \bibinfo {author} {\bibfnamefont {A.}~\bibnamefont {Miranowicz}},
  \bibinfo {author} {\bibfnamefont {S.}~\bibnamefont {{De Liberato}}}, \bibinfo
  {author} {\bibfnamefont {S.}~\bibnamefont {Savasta}}, \ and\ \bibinfo
  {author} {\bibfnamefont {F.}~\bibnamefont {Nori}},\ }\bibfield  {title}
  {\enquote {\bibinfo {title} {{Ultrastrong coupling between light and
  matter}},}\ }\href@noop {} {\bibfield  {journal} {\bibinfo  {journal} {Nat.
  Rev. Phys.}\ }\textbf {\bibinfo {volume} {1}},\ \bibinfo {pages} {19--40}
  (\bibinfo {year} {2019})}\BibitemShut {NoStop}%
\bibitem [{\citenamefont {Devoret}\ and\ \citenamefont
  {Martinis}(2004)}]{Devoret2004}%
  \BibitemOpen
  \bibfield  {author} {\bibinfo {author} {\bibfnamefont {M.~H.}\ \bibnamefont
  {Devoret}}\ and\ \bibinfo {author} {\bibfnamefont {J.~M.}\ \bibnamefont
  {Martinis}},\ }\bibfield  {title} {\enquote {\bibinfo {title} {{Implementing
  Qubits with Superconducting Integrated Circuits}},}\ }\href {\doibase
  10.1007/s11128-004-3101-5} {\bibfield  {journal} {\bibinfo  {journal} {Quant.
  Inf. Proc.}\ }\textbf {\bibinfo {volume} {3}},\ \bibinfo {pages} {163--203}
  (\bibinfo {year} {2004})}\BibitemShut {NoStop}%
\bibitem [{\citenamefont {Leggett}\ \emph {et~al.}(1987)\citenamefont
  {Leggett}, \citenamefont {Chakravarty}, \citenamefont {Dorsey}, \citenamefont
  {Fisher}, \citenamefont {Garg},\ and\ \citenamefont {Zwerger}}]{Leggett1987}%
  \BibitemOpen
  \bibfield  {author} {\bibinfo {author} {\bibfnamefont {A.~J.}\ \bibnamefont
  {Leggett}}, \bibinfo {author} {\bibfnamefont {S.}~\bibnamefont
  {Chakravarty}}, \bibinfo {author} {\bibfnamefont {A.~T.}\ \bibnamefont
  {Dorsey}}, \bibinfo {author} {\bibfnamefont {M.~P.~A.}\ \bibnamefont
  {Fisher}}, \bibinfo {author} {\bibfnamefont {A.}~\bibnamefont {Garg}}, \ and\
  \bibinfo {author} {\bibfnamefont {W.}~\bibnamefont {Zwerger}},\ }\bibfield
  {title} {\enquote {\bibinfo {title} {{Dynamics of the dissipative two-state
  system}},}\ }\href {\doibase 10.1103/RevModPhys.59.1} {\bibfield  {journal}
  {\bibinfo  {journal} {Rev. Mod. Phys.}\ }\textbf {\bibinfo {volume} {59}},\
  \bibinfo {pages} {1--85} (\bibinfo {year} {1987})}\BibitemShut {NoStop}%
\bibitem [{\citenamefont {Weiss}(2012, 4th ed.)}]{Weiss2012}%
  \BibitemOpen
  \bibfield  {author} {\bibinfo {author} {\bibfnamefont {U.}~\bibnamefont
  {Weiss}},\ }\href@noop {} {\emph {\bibinfo {title} {{Quantum Dissipative
  Systems}}}}\ (\bibinfo  {publisher} {World Scientific, Singapore},\ \bibinfo
  {year} {2012, 4th ed.})\BibitemShut {NoStop}%
\bibitem [{\citenamefont {Breuer}\ and\ \citenamefont
  {Petruccione}(2002)}]{Petruccione2002}%
  \BibitemOpen
  \bibfield  {author} {\bibinfo {author} {\bibfnamefont {H.~P.}\ \bibnamefont
  {Breuer}}\ and\ \bibinfo {author} {\bibfnamefont {F.}~\bibnamefont
  {Petruccione}},\ }\href@noop {} {\emph {\bibinfo {title} {{The theory of open
  quantum systems}}}}\ (\bibinfo  {publisher} {Oxford University Press,
  Oxford},\ \bibinfo {year} {2002})\BibitemShut {NoStop}%
\bibitem [{\citenamefont {Forn-D{\'i}az}\ \emph {et~al.}(2017)\citenamefont
  {Forn-D{\'i}az}, \citenamefont {Garc{\'i}a-Ripoll}, \citenamefont
  {Peropadre}, \citenamefont {Orgiazzi}, \citenamefont {Yurtalan},
  \citenamefont {Belyansky}, \citenamefont {Wilson},\ and\ \citenamefont
  {Lupascu}}]{Forn-Diaz2017}%
  \BibitemOpen
  \bibfield  {author} {\bibinfo {author} {\bibfnamefont {P.}~\bibnamefont
  {Forn-D{\'i}az}}, \bibinfo {author} {\bibfnamefont {J.~J.}\ \bibnamefont
  {Garc{\'i}a-Ripoll}}, \bibinfo {author} {\bibfnamefont {B.}~\bibnamefont
  {Peropadre}}, \bibinfo {author} {\bibfnamefont {J.~L.}\ \bibnamefont
  {Orgiazzi}}, \bibinfo {author} {\bibfnamefont {M.~A.}\ \bibnamefont
  {Yurtalan}}, \bibinfo {author} {\bibfnamefont {R.}~\bibnamefont {Belyansky}},
  \bibinfo {author} {\bibfnamefont {C.~M.}\ \bibnamefont {Wilson}}, \ and\
  \bibinfo {author} {\bibfnamefont {A.}~\bibnamefont {Lupascu}},\ }\bibfield
  {title} {\enquote {\bibinfo {title} {{Ultrastrong coupling of a single
  artificial atom to an electromagnetic continuum in the nonperturbative
  regime}},}\ }\href {http://dx.doi.org/10.1038/nphys3905} {\bibfield
  {journal} {\bibinfo  {journal} {Nat. Phys.}\ }\textbf {\bibinfo {volume}
  {13}},\ \bibinfo {pages} {39--43} (\bibinfo {year} {2017})}\BibitemShut
  {NoStop}%
\bibitem [{\citenamefont {Magazz{\`u}}\ \emph {et~al.}(2018)\citenamefont
  {Magazz{\`u}}, \citenamefont {Forn-D{\'i}az}, \citenamefont {Belyansky},
  \citenamefont {Orgiazzi}, \citenamefont {Yurtalan}, \citenamefont {Otto},
  \citenamefont {Lupascu}, \citenamefont {Wilson},\ and\ \citenamefont
  {Grifoni}}]{Magazzu2018}%
  \BibitemOpen
  \bibfield  {author} {\bibinfo {author} {\bibfnamefont {L.}~\bibnamefont
  {Magazz{\`u}}}, \bibinfo {author} {\bibfnamefont {P.}~\bibnamefont
  {Forn-D{\'i}az}}, \bibinfo {author} {\bibfnamefont {R.}~\bibnamefont
  {Belyansky}}, \bibinfo {author} {\bibfnamefont {J.-L.}\ \bibnamefont
  {Orgiazzi}}, \bibinfo {author} {\bibfnamefont {M.~A.}\ \bibnamefont
  {Yurtalan}}, \bibinfo {author} {\bibfnamefont {M.~R.}\ \bibnamefont {Otto}},
  \bibinfo {author} {\bibfnamefont {A.}~\bibnamefont {Lupascu}}, \bibinfo
  {author} {\bibfnamefont {C.~M.}\ \bibnamefont {Wilson}}, \ and\ \bibinfo
  {author} {\bibfnamefont {M.}~\bibnamefont {Grifoni}},\ }\bibfield  {title}
  {\enquote {\bibinfo {title} {{Probing the strongly driven spin-boson model in
  a superconducting quantum circuit}},}\ }\href {\doibase
  10.1038/s41467-018-03626-w} {\bibfield  {journal} {\bibinfo  {journal} {Nat.
  Commun.}\ }\textbf {\bibinfo {volume} {9}},\ \bibinfo {pages} {1403}
  (\bibinfo {year} {2018})}\BibitemShut {NoStop}%
\bibitem [{\citenamefont {Lepp{\"a}kangas}\ \emph {et~al.}(2018)\citenamefont
  {Lepp{\"a}kangas}, \citenamefont {Braum{\"u}ller}, \citenamefont {Hauck},
  \citenamefont {Reiner}, \citenamefont {Schwenk}, \citenamefont {Zanker},
  \citenamefont {Fritz}, \citenamefont {Ustinov}, \citenamefont {Weides},\ and\
  \citenamefont {Marthaler}}]{Ustinov2018}%
  \BibitemOpen
  \bibfield  {author} {\bibinfo {author} {\bibfnamefont {J.}~\bibnamefont
  {Lepp{\"a}kangas}}, \bibinfo {author} {\bibfnamefont {J.}~\bibnamefont
  {Braum{\"u}ller}}, \bibinfo {author} {\bibfnamefont {M.}~\bibnamefont
  {Hauck}}, \bibinfo {author} {\bibfnamefont {J.-M.}\ \bibnamefont {Reiner}},
  \bibinfo {author} {\bibfnamefont {I.}~\bibnamefont {Schwenk}}, \bibinfo
  {author} {\bibfnamefont {S.}~\bibnamefont {Zanker}}, \bibinfo {author}
  {\bibfnamefont {L.}~\bibnamefont {Fritz}}, \bibinfo {author} {\bibfnamefont
  {A.~V.}\ \bibnamefont {Ustinov}}, \bibinfo {author} {\bibfnamefont
  {M.}~\bibnamefont {Weides}}, \ and\ \bibinfo {author} {\bibfnamefont
  {M.}~\bibnamefont {Marthaler}},\ }\bibfield  {title} {\enquote {\bibinfo
  {title} {{Quantum simulation of the spin-boson model with a microwave
  circuit}},}\ }\href {\doibase 10.1103/PhysRevA.97.052321} {\bibfield
  {journal} {\bibinfo  {journal} {Phys. Rev. A}\ }\textbf {\bibinfo {volume}
  {97}},\ \bibinfo {pages} {052321} (\bibinfo {year} {2018})}\BibitemShut
  {NoStop}%
\bibitem [{\citenamefont {Javier}\ \emph {et~al.}(2019)\citenamefont {Javier},
  \citenamefont {S{\'e}bastien}, \citenamefont {Gheeraert}, \citenamefont
  {Dassonneville}, \citenamefont {Planat}, \citenamefont {Foroughi},
  \citenamefont {Krupko}, \citenamefont {Buisson}, \citenamefont {Naud},
  \citenamefont {Hasch-Guichard}, \citenamefont {Florens}, \citenamefont
  {Snyman},\ and\ \citenamefont {Roch}}]{Roch2019}%
  \BibitemOpen
  \bibfield  {author} {\bibinfo {author} {\bibfnamefont {P.~M.}\ \bibnamefont
  {Javier}}, \bibinfo {author} {\bibfnamefont {L.}~\bibnamefont
  {S{\'e}bastien}}, \bibinfo {author} {\bibfnamefont {N.}~\bibnamefont
  {Gheeraert}}, \bibinfo {author} {\bibfnamefont {R.}~\bibnamefont
  {Dassonneville}}, \bibinfo {author} {\bibfnamefont {L.}~\bibnamefont
  {Planat}}, \bibinfo {author} {\bibfnamefont {F.}~\bibnamefont {Foroughi}},
  \bibinfo {author} {\bibfnamefont {Y.}~\bibnamefont {Krupko}}, \bibinfo
  {author} {\bibfnamefont {O.}~\bibnamefont {Buisson}}, \bibinfo {author}
  {\bibfnamefont {C.}~\bibnamefont {Naud}}, \bibinfo {author} {\bibfnamefont
  {W.}~\bibnamefont {Hasch-Guichard}}, \bibinfo {author} {\bibfnamefont
  {S.}~\bibnamefont {Florens}}, \bibinfo {author} {\bibfnamefont
  {I.}~\bibnamefont {Snyman}}, \ and\ \bibinfo {author} {\bibfnamefont
  {N.}~\bibnamefont {Roch}},\ }\bibfield  {title} {\enquote {\bibinfo {title}
  {{A tunable Josephson platform to explore many-body quantum optics in
  circuit-QED}},}\ }\href@noop {} {\bibfield  {journal} {\bibinfo  {journal}
  {npj Quantum Inf.}\ }\textbf {\bibinfo {volume} {5}},\ \bibinfo {pages} {19}
  (\bibinfo {year} {2019})}\BibitemShut {NoStop}%
\bibitem [{\citenamefont {Kuzmin}\ \emph {et~al.}(2019)\citenamefont {Kuzmin},
  \citenamefont {Mehta}, \citenamefont {Grabon}, \citenamefont {Mencia},\ and\
  \citenamefont {Manucharyan}}]{Kuzmin2019}%
  \BibitemOpen
  \bibfield  {author} {\bibinfo {author} {\bibfnamefont {R.}~\bibnamefont
  {Kuzmin}}, \bibinfo {author} {\bibfnamefont {N}~\bibnamefont {Mehta}},
  \bibinfo {author} {\bibfnamefont {N.}~\bibnamefont {Grabon}}, \bibinfo
  {author} {\bibfnamefont {R.}~\bibnamefont {Mencia}}, \ and\ \bibinfo {author}
  {\bibfnamefont {V.~E.}\ \bibnamefont {Manucharyan}},\ }\bibfield  {title}
  {\enquote {\bibinfo {title} {{Superstrong coupling in circuit quantum
  electrodynamics}},}\ }\href@noop {} {\bibfield  {journal} {\bibinfo
  {journal} {npj Quantum Inf.}\ }\textbf {\bibinfo {volume} {5}},\ \bibinfo
  {pages} {20} (\bibinfo {year} {2019})}\BibitemShut {NoStop}%
\bibitem [{\citenamefont {Rossatto}\ \emph {et~al.}(2017)\citenamefont
  {Rossatto}, \citenamefont {Villas-B{\^o}as}, \citenamefont {Sanz},\ and\
  \citenamefont {Solano}}]{Rossatto2017}%
  \BibitemOpen
  \bibfield  {author} {\bibinfo {author} {\bibfnamefont {D.~Z.}\ \bibnamefont
  {Rossatto}}, \bibinfo {author} {\bibfnamefont {C.~J.}\ \bibnamefont
  {Villas-B{\^o}as}}, \bibinfo {author} {\bibfnamefont {M.}~\bibnamefont
  {Sanz}}, \ and\ \bibinfo {author} {\bibfnamefont {E.}~\bibnamefont
  {Solano}},\ }\bibfield  {title} {\enquote {\bibinfo {title} {Spectral
  classification of coupling regimes in the quantum Rabi model},}\ }\href
  {\doibase 10.1103/PhysRevA.96.013849} {\bibfield  {journal} {\bibinfo
  {journal} {Phys. Rev. A}\ }\textbf {\bibinfo {volume} {96}},\ \bibinfo
  {pages} {013849} (\bibinfo {year} {2017})}\BibitemShut {NoStop}%
\bibitem [{\citenamefont {Garziano}\ \emph {et~al.}(2015)\citenamefont
  {Garziano}, \citenamefont {Stassi}, \citenamefont {Macr{\`\i}}, \citenamefont
  {Kockum}, \citenamefont {Savasta},\ and\ \citenamefont
  {Nori}}]{Garziano2015}%
  \BibitemOpen
  \bibfield  {author} {\bibinfo {author} {\bibfnamefont {L.}~\bibnamefont
  {Garziano}}, \bibinfo {author} {\bibfnamefont {R.}~\bibnamefont {Stassi}},
  \bibinfo {author} {\bibfnamefont {V.}~\bibnamefont {Macr{\`\i}}}, \bibinfo
  {author} {\bibfnamefont {A.~F.}\ \bibnamefont {Kockum}}, \bibinfo {author}
  {\bibfnamefont {S.}~\bibnamefont {Savasta}}, \ and\ \bibinfo {author}
  {\bibfnamefont {F.}~\bibnamefont {Nori}},\ }\bibfield  {title} {\enquote
  {\bibinfo {title} {Multiphoton quantum Rabi oscillations in ultrastrong
  cavity QED},}\ }\href {\doibase 10.1103/PhysRevA.92.063830} {\bibfield
  {journal} {\bibinfo  {journal} {Phys. Rev. A}\ }\textbf {\bibinfo {volume}
  {92}},\ \bibinfo {pages} {063830} (\bibinfo {year} {2015})}\BibitemShut
  {NoStop}%
\bibitem [{\citenamefont {D{\'\i}az-Camacho}\ \emph {et~al.}(2016)\citenamefont
  {D{\'\i}az-Camacho}, \citenamefont {Bermudez},\ and\ \citenamefont
  {Garc{\'\i}a-Ripoll}}]{DiazCamacho2016}%
  \BibitemOpen
  \bibfield  {author} {\bibinfo {author} {\bibfnamefont {G.}~\bibnamefont
  {D{\'\i}az-Camacho}}, \bibinfo {author} {\bibfnamefont {A.}~\bibnamefont
  {Bermudez}}, \ and\ \bibinfo {author} {\bibfnamefont {J.~J.}\ \bibnamefont
  {Garc{\'\i}a-Ripoll}},\ }\bibfield  {title} {\enquote {\bibinfo {title}
  {Dynamical polaron ansatz: A theoretical tool for the ultrastrong-coupling
  regime of circuit QED},}\ }\href {\doibase 10.1103/PhysRevA.93.043843}
  {\bibfield  {journal} {\bibinfo  {journal} {Phys. Rev. Lett.}\ }\textbf
  {\bibinfo {volume} {93}},\ \bibinfo {pages} {043843} (\bibinfo {year}
  {2016})}\BibitemShut {NoStop}%
\bibitem [{\citenamefont {Armata}\ \emph {et~al.}(2017)\citenamefont {Armata},
  \citenamefont {Calajo}, \citenamefont {Jaako}, \citenamefont {Kim},\ and\
  \citenamefont {Rabl}}]{Armata2017}%
  \BibitemOpen
  \bibfield  {author} {\bibinfo {author} {\bibfnamefont {F.}~\bibnamefont
  {Armata}}, \bibinfo {author} {\bibfnamefont {G.}~\bibnamefont {Calajo}},
  \bibinfo {author} {\bibfnamefont {T.}~\bibnamefont {Jaako}}, \bibinfo
  {author} {\bibfnamefont {M.~S.}\ \bibnamefont {Kim}}, \ and\ \bibinfo
  {author} {\bibfnamefont {P.}~\bibnamefont {Rabl}},\ }\bibfield  {title}
  {\enquote {\bibinfo {title} {Harvesting multiqubit entanglement from
  ultrastrong interactions in circuit quantum electrodynamics},}\ }\href
  {\doibase 10.1103/PhysRevLett.119.183602} {\bibfield  {journal} {\bibinfo
  {journal} {Phys. Rev. Lett.}\ }\textbf {\bibinfo {volume} {119}},\ \bibinfo
  {pages} {183602} (\bibinfo {year} {2017})}\BibitemShut {NoStop}%
\bibitem [{\citenamefont {{De Bernardis}}\ \emph {et~al.}(2018)\citenamefont
  {{De Bernardis}}, \citenamefont {Pilar}, \citenamefont {Jaako}, \citenamefont
  {{De Liberato}},\ and\ \citenamefont {Rabl}}]{DeBernardis2018}%
  \BibitemOpen
  \bibfield  {author} {\bibinfo {author} {\bibfnamefont {D.}~\bibnamefont {{De
  Bernardis}}}, \bibinfo {author} {\bibfnamefont {P.}~\bibnamefont {Pilar}},
  \bibinfo {author} {\bibfnamefont {T.}~\bibnamefont {Jaako}}, \bibinfo
  {author} {\bibfnamefont {S.}~\bibnamefont {{De Liberato}}}, \ and\ \bibinfo
  {author} {\bibfnamefont {P.}~\bibnamefont {Rabl}},\ }\bibfield  {title}
  {\enquote {\bibinfo {title} {{Breakdown of gauge invariance in
  ultrastrong-coupling cavity QED}},}\ }\href@noop {} {\bibfield  {journal}
  {\bibinfo  {journal} {Phys. Rev. A}\ }\textbf {\bibinfo {volume} {98}},\
  \bibinfo {pages} {053819} (\bibinfo {year} {2018})}\BibitemShut {NoStop}%
\bibitem [{\citenamefont {Di~Stefano}\ \emph {et~al.}(2019)\citenamefont
  {Di~Stefano}, \citenamefont {Settineri}, \citenamefont {Macr{\`\i}},
  \citenamefont {Garziano}, \citenamefont {Stassi}, \citenamefont {Savasta},\
  and\ \citenamefont {Nori}}]{DiStefano2019}%
  \BibitemOpen
  \bibfield  {author} {\bibinfo {author} {\bibfnamefont {O.}~\bibnamefont
  {Di~Stefano}}, \bibinfo {author} {\bibfnamefont {A.}~\bibnamefont
  {Settineri}}, \bibinfo {author} {\bibfnamefont {V.}~\bibnamefont
  {Macr{\`\i}}}, \bibinfo {author} {\bibfnamefont {L.}~\bibnamefont
  {Garziano}}, \bibinfo {author} {\bibfnamefont {R.}~\bibnamefont {Stassi}},
  \bibinfo {author} {\bibfnamefont {S.}~\bibnamefont {Savasta}}, \ and\
  \bibinfo {author} {\bibfnamefont {F.}~\bibnamefont {Nori}},\ }\bibfield
  {title} {\enquote {\bibinfo {title} {Resolution of gauge ambiguities in
  ultrastrong-coupling cavity quantum electrodynamics},}\ }\href {\doibase
  10.1038/s41567-019-0534-4} {\bibfield  {journal} {\bibinfo  {journal} {Nat.
  Phys.}\ }\textbf {\bibinfo {volume} {15}},\ \bibinfo {pages} {803--808}
  (\bibinfo {year} {2019})}\BibitemShut {NoStop}%
\bibitem [{\citenamefont {Beaudoin}\ \emph {et~al.}(2011)\citenamefont
  {Beaudoin}, \citenamefont {Gambetta},\ and\ \citenamefont
  {Blais}}]{Blais2011}%
  \BibitemOpen
  \bibfield  {author} {\bibinfo {author} {\bibfnamefont {F.}~\bibnamefont
  {Beaudoin}}, \bibinfo {author} {\bibfnamefont {J.~M.}\ \bibnamefont
  {Gambetta}}, \ and\ \bibinfo {author} {\bibfnamefont {A.}~\bibnamefont
  {Blais}},\ }\bibfield  {title} {\enquote {\bibinfo {title} {{Dissipation and
  ultrastrong coupling in circuit QED}},}\ }\href@noop {} {\bibfield  {journal}
  {\bibinfo  {journal} {Phys. Rev. A}\ }\textbf {\bibinfo {volume} {84}},\
  \bibinfo {pages} {043832} (\bibinfo {year} {2011})}\BibitemShut {NoStop}%
\bibitem [{\citenamefont {Chiorescu}\ \emph {et~al.}(2004)\citenamefont
  {Chiorescu}, \citenamefont {Bertet}, \citenamefont {Semba}, \citenamefont
  {Nakamura}, \citenamefont {Harmans},\ and\ \citenamefont
  {Mooij}}]{Chiorescu2004}%
  \BibitemOpen
  \bibfield  {author} {\bibinfo {author} {\bibfnamefont {I.}~\bibnamefont
  {Chiorescu}}, \bibinfo {author} {\bibfnamefont {P.}~\bibnamefont {Bertet}},
  \bibinfo {author} {\bibfnamefont {K.}~\bibnamefont {Semba}}, \bibinfo
  {author} {\bibfnamefont {Y.}~\bibnamefont {Nakamura}}, \bibinfo {author}
  {\bibfnamefont {C.~J. P.~M.}\ \bibnamefont {Harmans}}, \ and\ \bibinfo
  {author} {\bibfnamefont {J.~E.}\ \bibnamefont {Mooij}},\ }\bibfield  {title}
  {\enquote {\bibinfo {title} {{Coherent dynamics of a flux qubit coupled to a
  harmonic oscillator}},}\ }\href@noop {} {\bibfield  {journal} {\bibinfo
  {journal} {Nature}\ }\textbf {\bibinfo {volume} {431}},\ \bibinfo {pages}
  {159--162} (\bibinfo {year} {2004})}\BibitemShut {NoStop}%
\bibitem [{\citenamefont {Thorwart}\ \emph {et~al.}(2004)\citenamefont
  {Thorwart}, \citenamefont {Paladino},\ and\ \citenamefont
  {Grifoni}}]{Thorwart2004}%
  \BibitemOpen
  \bibfield  {author} {\bibinfo {author} {\bibfnamefont {M.}~\bibnamefont
  {Thorwart}}, \bibinfo {author} {\bibfnamefont {E.}~\bibnamefont {Paladino}},
  \ and\ \bibinfo {author} {\bibfnamefont {M.}~\bibnamefont {Grifoni}},\
  }\bibfield  {title} {\enquote {\bibinfo {title} {{Dynamics of the spin-boson
  model with a structured environment}},}\ }\href@noop {} {\bibfield  {journal}
  {\bibinfo  {journal} {Chem. Phys}\ }\textbf {\bibinfo {volume} {296}},\
  \bibinfo {pages} {333--344} (\bibinfo {year} {2004})}\BibitemShut {NoStop}%
\bibitem [{\citenamefont {Johansson}\ \emph {et~al.}(2006)\citenamefont
  {Johansson}, \citenamefont {Saito}, \citenamefont {Meno}, \citenamefont
  {Nakano}, \citenamefont {Ueda}, \citenamefont {Semba},\ and\ \citenamefont
  {Takayanagi}}]{Johansson2006}%
  \BibitemOpen
  \bibfield  {author} {\bibinfo {author} {\bibfnamefont {J.}~\bibnamefont
  {Johansson}}, \bibinfo {author} {\bibfnamefont {S.}~\bibnamefont {Saito}},
  \bibinfo {author} {\bibfnamefont {T.}~\bibnamefont {Meno}}, \bibinfo {author}
  {\bibfnamefont {H.}~\bibnamefont {Nakano}}, \bibinfo {author} {\bibfnamefont
  {M.}~\bibnamefont {Ueda}}, \bibinfo {author} {\bibfnamefont {K.}~\bibnamefont
  {Semba}}, \ and\ \bibinfo {author} {\bibfnamefont {H.}~\bibnamefont
  {Takayanagi}},\ }\bibfield  {title} {\enquote {\bibinfo {title} {{Vacuum Rabi
  Oscillations in a Macroscopic Superconducting Qubit $LC$ Oscillator
  System}},}\ }\href@noop {} {\bibfield  {journal} {\bibinfo  {journal} {Phys.
  Rev. Lett.}\ }\textbf {\bibinfo {volume} {96}},\ \bibinfo {pages} {127006}
  (\bibinfo {year} {2006})}\BibitemShut {NoStop}%
\bibitem [{\citenamefont {{A. Ronzani}}\ \emph {et~al.}(2018)\citenamefont {{A.
  Ronzani}}, \citenamefont {{B. Karimi}}, \citenamefont {{J. Senior}},
  \citenamefont {{Y.-C. Chang}}, \citenamefont {{J. T. Peltonen}},
  \citenamefont {{C.D. Chen}},\ and\ \citenamefont {{J. P.
  Pekola}}}]{Pekola2018}%
  \BibitemOpen
  \bibfield  {author} {\bibinfo {author} {\bibnamefont {{A. Ronzani}}},
  \bibinfo {author} {\bibnamefont {{B. Karimi}}}, \bibinfo {author}
  {\bibnamefont {{J. Senior}}}, \bibinfo {author} {\bibnamefont {{Y.-C.
  Chang}}}, \bibinfo {author} {\bibnamefont {{J. T. Peltonen}}}, \bibinfo
  {author} {\bibnamefont {{C.D. Chen}}}, \ and\ \bibinfo {author} {\bibnamefont
  {{J. P. Pekola}}},\ }\bibfield  {title} {\enquote {\bibinfo {title} {{Tunable
  photonic heat transport in a quantum heat valve}},}\ }\href {\doibase
  10.1038/s41567-018-0199-4} {\bibfield  {journal} {\bibinfo  {journal} {Nat.
  Phys.}\ }\textbf {\bibinfo {volume} {14}},\ \bibinfo {pages} {991--995}
  (\bibinfo {year} {2018})}\BibitemShut {NoStop}%
\bibitem [{\citenamefont {Lolli}\ \emph {et~al.}(2015)\citenamefont {Lolli},
  \citenamefont {Baksic}, \citenamefont {Nagy}, \citenamefont {Manucharyan},\
  and\ \citenamefont {Ciuti}}]{Lolli2015}%
  \BibitemOpen
  \bibfield  {author} {\bibinfo {author} {\bibfnamefont {J.}~\bibnamefont
  {Lolli}}, \bibinfo {author} {\bibfnamefont {A.}~\bibnamefont {Baksic}},
  \bibinfo {author} {\bibfnamefont {D.}~\bibnamefont {Nagy}}, \bibinfo {author}
  {\bibfnamefont {V.~E.}\ \bibnamefont {Manucharyan}}, \ and\ \bibinfo {author}
  {\bibfnamefont {C.}~\bibnamefont {Ciuti}},\ }\bibfield  {title} {\enquote
  {\bibinfo {title} {Ancillary qubit spectroscopy of vacua in cavity and
  circuit quantum electrodynamics},}\ }\href {\doibase
  10.1103/PhysRevLett.114.183601} {\bibfield  {journal} {\bibinfo  {journal}
  {Phys. Rev. Lett.}\ }\textbf {\bibinfo {volume} {114}},\ \bibinfo {pages}
  {183601} (\bibinfo {year} {2015})}\BibitemShut {NoStop}%
\bibitem [{\citenamefont {{G. Falci}}\ \emph {et~al.}(2018)\citenamefont {{G. Falci}}, \citenamefont {{A. Ridolfo}},
  \citenamefont {{P. G. Di Stefano}},\ and\ \citenamefont {{E. Paladino}}}]{Falci2019}%
  \BibitemOpen
  \bibfield  {author} {\bibinfo {author} {\bibnamefont {{G. Falci}}},
  \bibinfo {author} {\bibnamefont {{A. Ridolfo}}}, \bibinfo {author}
  {\bibnamefont {{P. G. Di Stefano}}}, \ and\ \bibinfo {author} {\bibnamefont
  {{E. Paladino}}},\ }\bibfield  {title} {\enquote {\bibinfo {title} {{Ultrastrong coupling probed by coherent population transfer}},}\ }\href {\doibase
 10.1038/s41598-019-45187-y} {\bibfield  {journal} {\bibinfo  {journal} {Sci.
 Rep.}\ }\textbf {\bibinfo {volume} {9}},\ \bibinfo {pages} {9249}
  (\bibinfo {year} {2019})}\BibitemShut {NoStop}%
\bibitem [{\citenamefont {{A. Ridolfo}}\ \emph {et~al.}(2018) \citenamefont {{A. Ridolfo}},\citenamefont {{G. Falci}},
  \citenamefont {{F. M. D. Pellegrino}},  \citenamefont {{G. D. Maccarrone}},\ and\ \citenamefont {{E. Paladino}}}]{Ridolfo2019}%
  \BibitemOpen
  \bibfield  {author} {\bibinfo {author} {\bibnamefont {{A. Ridolfo}}},
  \bibinfo {author} {\bibnamefont {{G. Falci}}}, \bibinfo {author}
  {\bibnamefont {{F. M. D. Pellegrino}}}, \bibinfo {author}
  {\bibnamefont {{G. D. Maccarrone}}}, \ and\ \bibinfo {author} {\bibnamefont
  {{E. Paladino}}},\ }\bibfield  {title} {\enquote {\bibinfo {title} {{Photon pair production by STIRAP in ultrastrongly coupled matter-radiation systems}},}\ }\href {\doibase
 10.1140/epjst/e2018-800076-1} {\bibfield  {journal} {\bibinfo  {journal} {Eur. Phys. J. Spec. Top.}\ }\textbf {\bibinfo {volume} {227}},\ \bibinfo {pages} {2183}
  (\bibinfo {year} {2019})}\BibitemShut {NoStop}%
\bibitem [{\citenamefont {Garg}\ \emph {et~al.}(1985)\citenamefont {Garg},
  \citenamefont {Onuchic},\ and\ \citenamefont {Ambegaokar}}]{Garg1985}%
  \BibitemOpen
  \bibfield  {author} {\bibinfo {author} {\bibfnamefont {A.}~\bibnamefont
  {Garg}}, \bibinfo {author} {\bibfnamefont {J.~N.}\ \bibnamefont {Onuchic}}, \
  and\ \bibinfo {author} {\bibfnamefont {V.}~\bibnamefont {Ambegaokar}},\
  }\bibfield  {title} {\enquote {\bibinfo {title} {{Effect of friction on
  electron transfer in biomolecules}},}\ }\href {\doibase 10.1063/1.449017}
  {\bibfield  {journal} {\bibinfo  {journal} {J. Chem. Phys.}\ }\textbf
  {\bibinfo {volume} {83}},\ \bibinfo {pages} {4491--4503} (\bibinfo {year}
  {1985})}\BibitemShut {NoStop}%
\bibitem [{\citenamefont {{Van Vleck}}(1929)}]{VanVleck1929}%
  \BibitemOpen
  \bibfield  {author} {\bibinfo {author} {\bibfnamefont {J.~H.}\ \bibnamefont
  {{Van Vleck}}},\ }\bibfield  {title} {\enquote {\bibinfo {title} {{On
  $\ensuremath{\sigma}$-Type Doubling and Electron Spin in the Spectra of
  Diatomic Molecules}},}\ }\href@noop {} {\bibfield  {journal} {\bibinfo
  {journal} {Phys. Rev.}\ }\textbf {\bibinfo {volume} {33}},\ \bibinfo {pages}
  {467--506} (\bibinfo {year} {1929})}\BibitemShut {NoStop}%
\bibitem [{\citenamefont {Cohen-Tannoudji}\ \emph {et~al.}(1998)\citenamefont
  {Cohen-Tannoudji}, \citenamefont {Dupont-Roc},\ and\ \citenamefont
  {Grynberg}}]{CohenTannoudji1998}%
  \BibitemOpen
  \bibfield  {author} {\bibinfo {author} {\bibfnamefont {C.}~\bibnamefont
  {Cohen-Tannoudji}}, \bibinfo {author} {\bibfnamefont {J.}~\bibnamefont
  {Dupont-Roc}}, \ and\ \bibinfo {author} {\bibfnamefont {G.}~\bibnamefont
  {Grynberg}},\ }\href@noop {} {\emph {\bibinfo {title} {{Atom-Photon
  Interactions: Basic Processes and Applications}}}}\ (\bibinfo  {publisher}
  {Wiley-VCH, New York},\ \bibinfo {year} {1998})\BibitemShut {NoStop}%
\bibitem [{\citenamefont {Hausinger}\ and\ \citenamefont
  {Grifoni}(2008)}]{Hausinger2008}%
  \BibitemOpen
  \bibfield  {author} {\bibinfo {author} {\bibfnamefont {J.}~\bibnamefont
  {Hausinger}}\ and\ \bibinfo {author} {\bibfnamefont {M.}~\bibnamefont
  {Grifoni}},\ }\bibfield  {title} {\enquote {\bibinfo {title} {{Dissipative
  dynamics of a biased qubit coupled to a harmonic oscillator: analytical
  results beyond the rotating wave approximation}},}\ }\href@noop {} {\bibfield
   {journal} {\bibinfo  {journal} {New J. Phys.}\ }\textbf {\bibinfo {volume}
  {10}},\ \bibinfo {pages} {115015} (\bibinfo {year} {2008})}\BibitemShut
  {NoStop}%
\bibitem [{\citenamefont {Wallraff}\ \emph {et~al.}(2004)\citenamefont
  {Wallraff}, \citenamefont {Schuster}, \citenamefont {Blais}, \citenamefont
  {Frunzio}, \citenamefont {Huang}, \citenamefont {Majer}, \citenamefont
  {Kumar}, \citenamefont {Girvin},\ and\ \citenamefont
  {Schoelkopf}}]{Wallraff2004}%
  \BibitemOpen
  \bibfield  {author} {\bibinfo {author} {\bibfnamefont {A.}~\bibnamefont
  {Wallraff}}, \bibinfo {author} {\bibfnamefont {D.~I.}\ \bibnamefont
  {Schuster}}, \bibinfo {author} {\bibfnamefont {A.}~\bibnamefont {Blais}},
  \bibinfo {author} {\bibfnamefont {L.}~\bibnamefont {Frunzio}}, \bibinfo
  {author} {\bibfnamefont {R.-S.}\ \bibnamefont {Huang}}, \bibinfo {author}
  {\bibfnamefont {J.}~\bibnamefont {Majer}}, \bibinfo {author} {\bibfnamefont
  {S.}~\bibnamefont {Kumar}}, \bibinfo {author} {\bibfnamefont {S.~M.}\
  \bibnamefont {Girvin}}, \ and\ \bibinfo {author} {\bibfnamefont {R.~J.}\
  \bibnamefont {Schoelkopf}},\ }\bibfield  {title} {\enquote {\bibinfo {title}
  {{Strong coupling of a single photon to a superconducting qubit using circuit
  quantum electrodynamics}},}\ }\href {\doibase 10.1038/nature02851} {\bibfield
   {journal} {\bibinfo  {journal} {Nature}\ }\textbf {\bibinfo {volume}
  {431}},\ \bibinfo {pages} {162--167} (\bibinfo {year} {2004})}\BibitemShut
  {NoStop}%
\bibitem [{\citenamefont {Fragner}\ \emph {et~al.}(2008)\citenamefont
  {Fragner}, \citenamefont {G{\"o}ppl}, \citenamefont {Fink}, \citenamefont
  {Baur}, \citenamefont {Bianchetti}, \citenamefont {Leek}, \citenamefont
  {Blais},\ and\ \citenamefont {Wallraff}}]{Wallraff2008}%
  \BibitemOpen
  \bibfield  {author} {\bibinfo {author} {\bibfnamefont {A.}~\bibnamefont
  {Fragner}}, \bibinfo {author} {\bibfnamefont {M.}~\bibnamefont {G{\"o}ppl}},
  \bibinfo {author} {\bibfnamefont {J.~M.}\ \bibnamefont {Fink}}, \bibinfo
  {author} {\bibfnamefont {M.}~\bibnamefont {Baur}}, \bibinfo {author}
  {\bibfnamefont {R.}~\bibnamefont {Bianchetti}}, \bibinfo {author}
  {\bibfnamefont {P.~J.}\ \bibnamefont {Leek}}, \bibinfo {author}
  {\bibfnamefont {A.}~\bibnamefont {Blais}}, \ and\ \bibinfo {author}
  {\bibfnamefont {A.}~\bibnamefont {Wallraff}},\ }\bibfield  {title} {\enquote
  {\bibinfo {title} {{Resolving Vacuum Fluctuations in an Electrical Circuit by
  Measuring the Lamb Shift}},}\ }\href {\doibase 10.1126/science.1164482}
  {\bibfield  {journal} {\bibinfo  {journal} {Science}\ }\textbf {\bibinfo
  {volume} {322}},\ \bibinfo {pages} {1357} (\bibinfo {year}
  {2008})}\BibitemShut {NoStop}%
\bibitem [{\citenamefont {Hausinger}\ and\ \citenamefont
  {Grifoni}(2010)}]{Hausinger2010PRA}%
  \BibitemOpen
  \bibfield  {author} {\bibinfo {author} {\bibfnamefont {J.}~\bibnamefont
  {Hausinger}}\ and\ \bibinfo {author} {\bibfnamefont {M.}~\bibnamefont
  {Grifoni}},\ }\bibfield  {title} {\enquote {\bibinfo {title}
  {{Qubit-oscillator system: An analytical treatment of the ultrastrong
  coupling regime}},}\ }\href@noop {} {\bibfield  {journal} {\bibinfo
  {journal} {Phys. Rev. A}\ }\textbf {\bibinfo {volume} {82}},\ \bibinfo
  {pages} {062320} (\bibinfo {year} {2010})}\BibitemShut {NoStop}%
\bibitem [{\citenamefont {Grifoni}\ and\ \citenamefont
  {H{\"a}nggi}(1998)}]{Grifoni1998}%
  \BibitemOpen
  \bibfield  {author} {\bibinfo {author} {\bibfnamefont {M.}~\bibnamefont
  {Grifoni}}\ and\ \bibinfo {author} {\bibfnamefont {P.}~\bibnamefont
  {H{\"a}nggi}},\ }\bibfield  {title} {\enquote {\bibinfo {title} {{Driven
  quantum tunneling}},}\ }\href@noop {} {\bibfield  {journal} {\bibinfo
  {journal} {Phys. Rep.}\ }\textbf {\bibinfo {volume} {304}},\ \bibinfo {pages}
  {229--354} (\bibinfo {year} {1998})}\BibitemShut {NoStop}%
\bibitem [{\citenamefont {Goorden}\ \emph {et~al.}(2004)\citenamefont
  {Goorden}, \citenamefont {Thorwart},\ and\ \citenamefont
  {Grifoni}}]{Goorden2004}%
  \BibitemOpen
  \bibfield  {author} {\bibinfo {author} {\bibfnamefont {M.~C.}\ \bibnamefont
  {Goorden}}, \bibinfo {author} {\bibfnamefont {M.}~\bibnamefont {Thorwart}}, \
  and\ \bibinfo {author} {\bibfnamefont {M.}~\bibnamefont {Grifoni}},\
  }\bibfield  {title} {\enquote {\bibinfo {title} {{Entanglement Spectroscopy
  of a Driven Solid-State Qubit and Its Detector}},}\ }\href {\doibase
  10.1103/PhysRevLett.93.267005} {\bibfield  {journal} {\bibinfo  {journal}
  {Phys. Rev. Lett.}\ }\textbf {\bibinfo {volume} {93}},\ \bibinfo {pages}
  {267005} (\bibinfo {year} {2004})}\BibitemShut {NoStop}%
\bibitem [{\citenamefont {Zueco}\ and\ \citenamefont
  {Garc{\'i}a-Ripoll}(2019)}]{Zueco2019}%
  \BibitemOpen
  \bibfield  {author} {\bibinfo {author} {\bibfnamefont {D.}~\bibnamefont
  {Zueco}}\ and\ \bibinfo {author} {\bibfnamefont {J.~J.}\ \bibnamefont
  {Garc{\'i}a-Ripoll}},\ }\bibfield  {title} {\enquote {\bibinfo {title}
  {{Ultrastrongly dissipative quantum Rabi model}},}\ }\href@noop {} {\bibfield
   {journal} {\bibinfo  {journal} {Phys. Rev. A}\ }\textbf {\bibinfo {volume}
  {99}},\ \bibinfo {pages} {013807} (\bibinfo {year} {2019})}\BibitemShut
  {NoStop}%
\bibitem [{\citenamefont {{ R. Martinazzo}}\ \emph {et~al.}(2011)\citenamefont
  {{ R. Martinazzo}}, \citenamefont {{B. Vacchini, B.}}, \citenamefont {{K. H.
  Hughes}},\ and\ \citenamefont {{I. Burghardt}}}]{Martinazzo2011}%
  \BibitemOpen
  \bibfield  {author} {\bibinfo {author} {\bibnamefont {{ R. Martinazzo}}},
  \bibinfo {author} {\bibnamefont {{B. Vacchini, B.}}}, \bibinfo {author}
  {\bibnamefont {{K. H. Hughes}}}, \ and\ \bibinfo {author} {\bibnamefont {{I.
  Burghardt}}},\ }\bibfield  {title} {\enquote {\bibinfo {title}
  {{Communication: Universal Markovian reduction of Brownian particle
  dynamics}},}\ }\href {\doibase 10.1063/1.3532408} {\bibfield  {journal}
  {\bibinfo  {journal} {J. Chem. Phys.}\ }\textbf {\bibinfo {volume} {134}},\
  \bibinfo {pages} {011101} (\bibinfo {year} {2011})}\BibitemShut {NoStop}%
\bibitem [{\citenamefont {Feynman}\ and\ \citenamefont {{Vernon
  Jr.}}(1963)}]{Feynman1963}%
  \BibitemOpen
  \bibfield  {author} {\bibinfo {author} {\bibfnamefont {R.~P}\ \bibnamefont
  {Feynman}}\ and\ \bibinfo {author} {\bibfnamefont {F.~L.}\ \bibnamefont
  {{Vernon Jr.}}},\ }\bibfield  {title} {\enquote {\bibinfo {title} {{The
  theory of a general quantum system interacting with a linear dissipative
  system}},}\ }\href {\doibase 10.1016/0003-4916(63)90068-X} {\bibfield
  {journal} {\bibinfo  {journal} {Ann. Phys. (N.Y.)}\ }\textbf {\bibinfo
  {volume} {24}},\ \bibinfo {pages} {118--173} (\bibinfo {year}
  {1963})}\BibitemShut {NoStop}%
\bibitem [{\citenamefont {H{\"a}nggi}\ and\ \citenamefont
  {Ingold}(2005)}]{Hanggi2005}%
  \BibitemOpen
  \bibfield  {author} {\bibinfo {author} {\bibfnamefont {P.}~\bibnamefont
  {H{\"a}nggi}}\ and\ \bibinfo {author} {\bibfnamefont {G.-L.}\ \bibnamefont
  {Ingold}},\ }\bibfield  {title} {\enquote {\bibinfo {title} {{Fundamental
  aspects of quantum Brownian motion}},}\ }\href {\doibase 10.1063/1.1853631}
  {\bibfield  {journal} {\bibinfo  {journal} {Chaos}\ }\textbf {\bibinfo
  {volume} {15}},\ \bibinfo {pages} {26105} (\bibinfo {year}
  {2005})}\BibitemShut {NoStop}%
\bibitem [{\citenamefont {{F. Nesi}}\ \emph {et~al.}(2007)\citenamefont {{F.
  Nesi}}, \citenamefont {{M. Grifoni}},\ and\ \citenamefont {{E.
  Paladino}}}]{Nesi2007}%
  \BibitemOpen
  \bibfield  {author} {\bibinfo {author} {\bibnamefont {{F. Nesi}}}, \bibinfo
  {author} {\bibnamefont {{M. Grifoni}}}, \ and\ \bibinfo {author}
  {\bibnamefont {{E. Paladino}}},\ }\bibfield  {title} {\enquote {\bibinfo
  {title} {{Dynamics of a qubit coupled to a broadened harmonic mode at finite
  detuning}},}\ }\href {\doibase 10.1088/1367-2630/9/9/316} {\bibfield
  {journal} {\bibinfo  {journal} {New J. Phys.}\ }\textbf {\bibinfo {volume}
  {9}},\ \bibinfo {pages} {316--316} (\bibinfo {year} {2007})}\BibitemShut
  {NoStop}%
\bibitem [{\citenamefont {Peropadre}\ \emph {et~al.}(2013)\citenamefont
  {Peropadre}, \citenamefont {Lindkvist}, \citenamefont {Hoi}, \citenamefont
  {Wilson}, \citenamefont {Garc{\'i}a-Ripoll}, \citenamefont {Delsing},\ and\
  \citenamefont {Johansson}}]{Peropadre2013}%
  \BibitemOpen
  \bibfield  {author} {\bibinfo {author} {\bibfnamefont {B.}~\bibnamefont
  {Peropadre}}, \bibinfo {author} {\bibfnamefont {J.}~\bibnamefont
  {Lindkvist}}, \bibinfo {author} {\bibfnamefont {I.-C.}\ \bibnamefont {Hoi}},
  \bibinfo {author} {\bibfnamefont {C.~M.}\ \bibnamefont {Wilson}}, \bibinfo
  {author} {\bibfnamefont {J.~J.}\ \bibnamefont {Garc{\'i}a-Ripoll}}, \bibinfo
  {author} {\bibfnamefont {P.}~\bibnamefont {Delsing}}, \ and\ \bibinfo
  {author} {\bibfnamefont {G.}~\bibnamefont {Johansson}},\ }\bibfield  {title}
  {\enquote {\bibinfo {title} {{Scattering of coherent states on a single
  artificial atom}},}\ }\href@noop {} {\bibfield  {journal} {\bibinfo
  {journal} {New J. Phys.}\ }\textbf {\bibinfo {volume} {15}},\ \bibinfo
  {pages} {035009} (\bibinfo {year} {2013})}\BibitemShut {NoStop}%
\bibitem [{\citenamefont {{U. Vool}}\ and\ \citenamefont {{M.
  Devoret}}(2017)}]{Vool2017}%
  \BibitemOpen
  \bibfield  {author} {\bibinfo {author} {\bibnamefont {{U. Vool}}}\ and\
  \bibinfo {author} {\bibnamefont {{M. Devoret}}},\ }\bibfield  {title}
  {\enquote {\bibinfo {title} {{Introduction to quantum electromagnetic
  circuits}},}\ }\href {\doibase 10.1002/cta.2359} {\bibfield  {journal}
  {\bibinfo  {journal} {Int. J. Circ. Theor. Appl.}\ }\textbf {\bibinfo
  {volume} {45}},\ \bibinfo {pages} {897--934} (\bibinfo {year}
  {2017})}\BibitemShut {NoStop}%
\bibitem [{\citenamefont {{M. Grifoni}}\ \emph {et~al.}(1995)\citenamefont {{M.
  Grifoni}}, \citenamefont {{M. Sassetti}}, \citenamefont {{P. H{\"a}nggi}},\
  and\ \citenamefont {{U. Weiss}}}]{Grifoni1995}%
  \BibitemOpen
  \bibfield  {author} {\bibinfo {author} {\bibnamefont {{M. Grifoni}}},
  \bibinfo {author} {\bibnamefont {{M. Sassetti}}}, \bibinfo {author}
  {\bibnamefont {{P. H{\"a}nggi}}}, \ and\ \bibinfo {author} {\bibnamefont {{U.
  Weiss}}},\ }\bibfield  {title} {\enquote {\bibinfo {title} {{Cooperative
  effects in the nonlinearly driven spin-boson system}},}\ }\href {\doibase
  10.1103/PhysRevE.52.3596} {\bibfield  {journal} {\bibinfo  {journal} {Phys.
  Rev. E}\ }\textbf {\bibinfo {volume} {52}},\ \bibinfo {pages} {3596--3607}
  (\bibinfo {year} {1995})}\BibitemShut {NoStop}%
\bibitem [{\citenamefont {Kleff}\ \emph {et~al.}(2004)\citenamefont {Kleff},
  \citenamefont {Kehrein},\ and\ \citenamefont {von Delft}}]{vonDelft2004}%
  \BibitemOpen
  \bibfield  {author} {\bibinfo {author} {\bibfnamefont {S.}~\bibnamefont
  {Kleff}}, \bibinfo {author} {\bibfnamefont {S.}~\bibnamefont {Kehrein}}, \
  and\ \bibinfo {author} {\bibfnamefont {J.}~\bibnamefont {von Delft}},\
  }\bibfield  {title} {\enquote {\bibinfo {title} {{Exploiting environmental
  resonances to enhance qubit quality factors}},}\ }\href@noop {} {\bibfield
  {journal} {\bibinfo  {journal} {Phys. Rev. B}\ }\textbf {\bibinfo {volume}
  {70}},\ \bibinfo {pages} {014516} (\bibinfo {year} {2004})}\BibitemShut
  {NoStop}%
\bibitem [{\citenamefont {Nicolin}\ and\ \citenamefont
  {Segal}(2011)}]{Segal2011}%
  \BibitemOpen
  \bibfield  {author} {\bibinfo {author} {\bibfnamefont {L.}~\bibnamefont
  {Nicolin}}\ and\ \bibinfo {author} {\bibfnamefont {D.}~\bibnamefont
  {Segal}},\ }\bibfield  {title} {\enquote {\bibinfo {title} {{Non-equilibrium
  spin-boson model: Counting statistics and the heat exchange fluctuation
  theorem}},}\ }\href {\doibase 10.1063/1.3655674} {\bibfield  {journal}
  {\bibinfo  {journal} {J. Chem. Phys.}\ }\textbf {\bibinfo {volume} {135}},\
  \bibinfo {pages} {164106} (\bibinfo {year} {2011})}\BibitemShut {NoStop}%
\bibitem [{\citenamefont {Boudjada}\ and\ \citenamefont
  {Segal}(2014)}]{Segal2014}%
  \BibitemOpen
  \bibfield  {author} {\bibinfo {author} {\bibfnamefont {N.}~\bibnamefont
  {Boudjada}}\ and\ \bibinfo {author} {\bibfnamefont {D.}~\bibnamefont
  {Segal}},\ }\bibfield  {title} {\enquote {\bibinfo {title} {{From Dissipative
  Dynamics to Studies of Heat Transfer at the Nanoscale: Analysis of the
  Spin-Boson Model}},}\ }\href {\doibase 10.1021/jp5091685} {\bibfield
  {journal} {\bibinfo  {journal} {J. Phys. Chem. A}\ }\textbf {\bibinfo
  {volume} {118}},\ \bibinfo {pages} {11323--11336} (\bibinfo {year}
  {2014})}\BibitemShut {NoStop}%
\bibitem [{\citenamefont {Segal}(2014)}]{Segal2014PRE}%
  \BibitemOpen
  \bibfield  {author} {\bibinfo {author} {\bibfnamefont {D.}~\bibnamefont
  {Segal}},\ }\bibfield  {title} {\enquote {\bibinfo {title} {{Heat transfer in
  the spin-boson model: A comparative study in the incoherent tunneling
  regime}},}\ }\href@noop {} {\bibfield  {journal} {\bibinfo  {journal} {Phys.
  Rev. E}\ }\textbf {\bibinfo {volume} {90}},\ \bibinfo {pages} {012148}
  (\bibinfo {year} {2014})}\BibitemShut {NoStop}%
\bibitem [{\citenamefont {Schmidt}\ \emph {et~al.}(2015)\citenamefont
  {Schmidt}, \citenamefont {Carusela}, \citenamefont {Pekola}, \citenamefont
  {Suomela},\ and\ \citenamefont {Ankerhold}}]{Ankerhold2015}%
  \BibitemOpen
  \bibfield  {author} {\bibinfo {author} {\bibfnamefont {R.}~\bibnamefont
  {Schmidt}}, \bibinfo {author} {\bibfnamefont {M.~F.}\ \bibnamefont
  {Carusela}}, \bibinfo {author} {\bibfnamefont {J.~P.}\ \bibnamefont
  {Pekola}}, \bibinfo {author} {\bibfnamefont {S.}~\bibnamefont {Suomela}}, \
  and\ \bibinfo {author} {\bibfnamefont {J.}~\bibnamefont {Ankerhold}},\
  }\bibfield  {title} {\enquote {\bibinfo {title} {{Work and heat for two-level
  systems in dissipative environments: Strong driving and non-Markovian
  dynamics}},}\ }\href {\doibase 10.1103/PhysRevB.91.224303} {\bibfield
  {journal} {\bibinfo  {journal} {Phys. Rev. B}\ }\textbf {\bibinfo {volume}
  {91}},\ \bibinfo {pages} {224303} (\bibinfo {year} {2015})}\BibitemShut
  {NoStop}%
\bibitem [{\citenamefont {Carrega}\ \emph {et~al.}(2015)\citenamefont
  {Carrega}, \citenamefont {Solinas}, \citenamefont {Braggio}, \citenamefont
  {Sassetti},\ and\ \citenamefont {Weiss}}]{Carrega2015}%
  \BibitemOpen
  \bibfield  {author} {\bibinfo {author} {\bibfnamefont {M.}~\bibnamefont
  {Carrega}}, \bibinfo {author} {\bibfnamefont {P.}~\bibnamefont {Solinas}},
  \bibinfo {author} {\bibfnamefont {A.}~\bibnamefont {Braggio}}, \bibinfo
  {author} {\bibfnamefont {M.}~\bibnamefont {Sassetti}}, \ and\ \bibinfo
  {author} {\bibfnamefont {U.}~\bibnamefont {Weiss}},\ }\bibfield  {title}
  {\enquote {\bibinfo {title} {{Functional integral approach to time-dependent
  heat exchange in open quantum systems: general method and applications}},}\
  }\href {\doibase 10.1088/1367-2630/17/4/045030} {\bibfield  {journal}
  {\bibinfo  {journal} {New J. Phys.}\ }\textbf {\bibinfo {volume} {17}},\
  \bibinfo {pages} {045030} (\bibinfo {year} {2015})}\BibitemShut {NoStop}%
\bibitem [{\citenamefont {Carrega}\ \emph {et~al.}(2016)\citenamefont
  {Carrega}, \citenamefont {Solinas}, \citenamefont {Sassetti},\ and\
  \citenamefont {Weiss}}]{Carrega2016}%
  \BibitemOpen
  \bibfield  {author} {\bibinfo {author} {\bibfnamefont {M.}~\bibnamefont
  {Carrega}}, \bibinfo {author} {\bibfnamefont {P.}~\bibnamefont {Solinas}},
  \bibinfo {author} {\bibfnamefont {M.}~\bibnamefont {Sassetti}}, \ and\
  \bibinfo {author} {\bibfnamefont {U.}~\bibnamefont {Weiss}},\ }\bibfield
  {title} {\enquote {\bibinfo {title} {{Energy Exchange in Driven Open Quantum
  Systems at Strong Coupling}},}\ }\href {\doibase
  10.1103/PhysRevLett.116.240403} {\bibfield  {journal} {\bibinfo  {journal}
  {Phys. Rev. Lett.}\ }\textbf {\bibinfo {volume} {116}},\ \bibinfo {pages}
  {240403} (\bibinfo {year} {2016})}\BibitemShut {NoStop}%
\bibitem [{\citenamefont {Wang}\ \emph {et~al.}(2017)\citenamefont {Wang},
  \citenamefont {Ren},\ and\ \citenamefont {Cao}}]{Wang2017}%
  \BibitemOpen
  \bibfield  {author} {\bibinfo {author} {\bibfnamefont {C.}~\bibnamefont
  {Wang}}, \bibinfo {author} {\bibfnamefont {J.}~\bibnamefont {Ren}}, \ and\
  \bibinfo {author} {\bibfnamefont {J.}~\bibnamefont {Cao}},\ }\bibfield
  {title} {\enquote {\bibinfo {title} {{Unifying quantum heat transfer in a
  nonequilibrium spin-boson model with full counting statistics}},}\ }\href
  {\doibase 10.1103/PhysRevA.95.023610} {\bibfield  {journal} {\bibinfo
  {journal} {Phys. Rev. A}\ }\textbf {\bibinfo {volume} {95}},\ \bibinfo
  {pages} {023610} (\bibinfo {year} {2017})}\BibitemShut {NoStop}%
\bibitem [{\citenamefont {Agarwalla}\ and\ \citenamefont
  {Segal}(2017)}]{Agarwalla2017}%
  \BibitemOpen
  \bibfield  {author} {\bibinfo {author} {\bibfnamefont {B.~K.}\ \bibnamefont
  {Agarwalla}}\ and\ \bibinfo {author} {\bibfnamefont {D.}~\bibnamefont
  {Segal}},\ }\bibfield  {title} {\enquote {\bibinfo {title} {{Energy current
  and its statistics in the nonequilibrium spin-boson model: Majorana fermion
  representation}},}\ }\href {\doibase 10.1088/1367-2630/aa6657} {\bibfield
  {journal} {\bibinfo  {journal} {New J. Phys.}\ }\textbf {\bibinfo {volume}
  {19}},\ \bibinfo {pages} {043030} (\bibinfo {year} {2017})}\BibitemShut
  {NoStop}%
\bibitem [{\citenamefont {Newman}\ \emph {et~al.}(2017)\citenamefont {Newman},
  \citenamefont {Mintert},\ and\ \citenamefont {Nazir}}]{Nazir2017}%
  \BibitemOpen
  \bibfield  {author} {\bibinfo {author} {\bibfnamefont {D.}~\bibnamefont
  {Newman}}, \bibinfo {author} {\bibfnamefont {F.}~\bibnamefont {Mintert}}, \
  and\ \bibinfo {author} {\bibfnamefont {A.}~\bibnamefont {Nazir}},\ }\bibfield
   {title} {\enquote {\bibinfo {title} {{Performance of a quantum heat engine
  at strong reservoir coupling}},}\ }\href@noop {} {\bibfield  {journal}
  {\bibinfo  {journal} {Phys. Rev. E}\ }\textbf {\bibinfo {volume} {95}},\
  \bibinfo {pages} {032139} (\bibinfo {year} {2017})}\BibitemShut {NoStop}%
\end{thebibliography}
%

\end{document}